\documentclass{IEEEtran}
\usepackage{cite}
\usepackage{amsmath,amssymb,amsfonts}
\usepackage{algorithmic}
\usepackage{graphicx}
\usepackage{textcomp}
\usepackage{fancyref}
\usepackage{booktabs}
\usepackage{algorithm}
\usepackage{algorithmic}
\usepackage{mathtools}
\usepackage{cleveref}
\usepackage{multirow}
\usepackage{amsthm}
\usepackage{enumerate}
\usepackage{changepage}   
\usepackage[utf8]{inputenc}
\usepackage[english]{babel}

\newcommand{\subVecgamma}{\gamma\hspace*{-.95ex}\gamma}

\graphicspath{{./}{./}}
\def\BibTeX{{\rm B\kern-.05em{\sc i\kern-.025em b}\kern-.08em
    T\kern-.1667em\lower.7ex\hbox{E}\kern-.125emX}}
\begin{document}
\title{Gridless DOA Estimation and Root-MUSIC for Non-Uniform Arrays}
\author{Mark Wagner, \IEEEmembership{Member, IEEE}, Yongsung Park, \IEEEmembership{Member, IEEE}, Peter Gerstoft, \IEEEmembership{Senior Member, IEEE} }

\maketitle
\begin{abstract}

The problem of gridless direction of arrival (DOA) estimation is addressed in the non-uniform array (NUA) case. Traditionally, gridless DOA estimation and root-MUSIC are only applicable for measurements from a uniform linear array (ULA). This is because the sample covariance matrix of ULA measurements has Toeplitz structure, and both algorithms are based on the Vandermonde decomposition of a Toeplitz matrix. The Vandermonde decomposition breaks a Toeplitz matrix into its harmonic components, from which the DOAs are estimated. First, we present the `irregular' Toeplitz matrix and irregular Vandermonde decomposition (IVD), which generalizes the Vandermonde decomposition to apply to a more general set of matrices. It is shown that the IVD is related to the MUSIC and root-MUSIC algorithms. Next, gridless DOA is generalized to the NUA case using IVD. The resulting non-convex optimization problem is solved using alternating projections (AP). A numerical analysis is performed on the AP based solution which shows that the generalization to NUAs has similar performance to traditional gridless DOA.
\end{abstract}

\today
\section{Introduction}
\label{sec:introduction}
\IEEEPARstart{E}{stimating} the direction of arrival (DOA) of one or more signals arriving at an array of sensors is an important topic in array signal processing and has a wide range of applications in radar, sonar, wireless communications, etc. Recently, the focus of DOA estimation has turned from classical subspace based DOA algorithms including MUSIC, root-MUSIC, and ESPRIT \cite{barabell1983improving, rao1989performance,pesavento2000unitary,van2004optimum} to newer compressive sensing based methods such as compressive DOA \cite{xenaki2014compressive, xenaki2015grid, gerstoft2015multiple}. 
Compressive methods have the advantage that they are high resolution (resolve nearby DOAs), and require only a single measurement snapshot.   

Early compressive DOA techniques approximated the measurements as a linear combination of a few array patterns from DOAs picked out of a grid of possible DOAs. However, this technique suffers from errors due to grid mismatch because true DOAs are not on a grid \cite{chi2011sensitivity}. As a response came the family of off-grid, or \textit{gridless} methods, which exploit sparsity in the \textit{atomic norm} of the measurements \cite{tang2013compressed,candes2014towards,yang2015gridless,park2019grid}. The atomic norm is the minimum number of ``atoms" from a manifold required to reconstruct a vector, thus gridless DOA is the continuous analog to gridded compressive DOA.

Gridless DOA is an application of the continuous compressed sensing (CCS) spectral estimation problem, which was introduced in \cite{tang2013compressed}. There have been many adaptations of CCS to related problems \cite{fazel2013hankel,chi2014compressive}, however, the atomic norm minimization formulation of gridless DOA for a line array is the focus of this work \cite{yang2016exact,yang2015enhancing}. There are other similar algorithms such as the enhanced matrix completion (EMaC) method \cite{chen2014robust} and gridless SPICE \cite{yang2015gridless} whose formulations are similar to gridless DOA. We refer the reader to \cite{YANG2018509} for a comprehensive review of CCS for DOA.

CCS involves solving a semi-definite programming problem (SDP) whose objective is to find the lowest rank Toeplitz matrix which can explain the measurements. The frequencies composing the signal can be recovered through Vandermonde decomposition of the optimal Toeplitz matrix \cite{pisarenko1973retrieval}. This decomposition is known to be unique when the Toeplitz matrix is rank deficient \cite{van2004optimum}. In the context of DOA estimation, the parameters of the Vandermonde matrices, known as harmonics, indicate the DOAs. The weakness of gridless methods is that they are limited to regularly sampled measurements that can only be taken from a uniform linear array (ULA).

There have been some recent efforts towards extending gridless DOA to NUAs \cite{yang2016vandermonde, semper2018grid, semper2019admm, raj2019super}. Many of these are based on array interpolation, where the manifold of a NUA is interpolated back to that of a ULA. This idea traces back to ``Fourier domain root-MUSIC" \cite{rubsamen2008direction}, which was used to extend the root-MUSIC algorithm to NUAs. There are many other interpolation based techniques for adapting older DOA algorithms to NUAs \cite{friedlander1992direction, friedlander1993root, doran1993coherent, weiss1995direction, gershman1997note, hyberg2004array, belloni2007DOA}. A drawback of these techniques is that the interpolation is inaccurate for the whole array field of view, and must be performed over many sectors of the array manifold.

This paper generalizes gridless DOA and root-MUSIC to NUAs by working directly on the NUA measurements (no interpolation). This is achieved through the \textit{irregular Toeplitz} and \textit{irregular Vandermonde} matrices. The word `irregular' refers to the irregular sampling of the wave field by a NUA. The irregular Toeplitz matrix can be decomposed into irregular Vandermonde components, similar to Toeplitz and Vandermonde matrices. This allows for both root-MUSIC and gridless DOA to be extended to NUAs. The DOAs are recovered through irregular Vandermonde decomposition (IVD) of the irregular Toeplitz matrix. 

The proposed extension of gridless DOA to NUAs is a rank minimization problem, which is non-convex \cite{chi2019nonconvex}. A non-convex optimization problem is typically substituted for its convex relaxation and solved using a general solver \cite{grant2008cvx}, such as the alternating directions method of multipliers (ADMM) \cite{boyd2004convex}. However, the proposed extension to NUAs cannot be easily cast to its convex relaxation.
Instead, the proposed solution is based on the alternating projections (AP) algorithm \cite[pg. 606]{dattorro2010convex}, which finds a point of intersection between two or more sets by iteratively projecting an initial estimate between the sets. 

The AP algorithm has been previously used to solve CCS problems \cite{condat2015cadzow}, and implementation of AP to gridless DOA for NUAs is formulated by projecting onto the irregular Toeplitz set. 
The use of AP for solving similar non-convex optimization problems gives promising results \cite{condat2015cadzow, cho2016fast, liu2017projected}. A comparison of AP based gridless DOA against several competing algorithms, including ADMM, reveals the AP solution has similar or superior performance. An analysis is provided of the algorithm's performance in challenging scenarios such as when the measurements are corrupted with noise or the DOAs are closely spaced.

The paper is structured as follows: Sec. \ref{sec:overview} contains an overview of gridless DOA and relevant concepts such as the Vandermonde decomposition and root-MUSIC algorithm. In Sec. \ref{sec:EVD}, the IVD is introduced, which allows any positive semi-definite (PSD) matrix to be decomposed into harmonics sampled at known irregular intervals (array sensor locations). The IVD allows for non-Toeplitz matrices to be decomposed into their harmonic components. The DOAs are given by the harmonics. 
In Sec. \ref{sec:contribution}, a modification of the gridless DOA problem is presented such that it can be applied to NUAs.
A solution to the reformulated gridless DOA problem based on the AP algorithm is proposed. In Sec. \ref{sec:Sim}, the proposed algorithm is tested on simulated measurements from both ULAs and NUAs. 
The results are for DOA estimation, but are also applicable to the broader CCS problem to irregularly sampled signals.

In the remainder of this paper, lowercase bold letters represent vectors and uppercase bold letters represent matrices. Below is a short list of notation that will be used.
\begin{itemize}
\item{\makebox[2cm] {$^{\mathsf{T}}$\hfill} Vector or matrix transpose.}
\item{\makebox[2cm]{$^*$\hfill} Complex conjugate.}
\item{\makebox[2cm]{$^{\mathsf{H}}$\hfill} Vector or matrix conjugate transpose.}
\item{\makebox[2cm]{$^{\dagger}$\hfill} Moore-Penrose pseudo-inverse.}
\item{\makebox[2cm]{$\angle$\hfill} Phase angle.}
\item{\makebox[2cm]{$\perp$\hfill} Orthogonal.}
\item{\makebox[2cm]{$\succeq$\hfill} Positive semi-definite.}
\item{\makebox[2cm]{$\|.\|_2$\hfill} Two norm.}
\item{\makebox[2cm]{$\|.\|_F$\hfill} Frobenious norm.}
\item{\makebox[2cm]{$\mathbb{E}$\hfill} Expected value.}
\item{\makebox[2cm]{$\mathrm{Tr}$\hfill} Matrix trace.}
\item{\makebox[2cm]{$\mathrm{diag}(\mathbf{X})$\hfill} Diagonal elements of matrix $\mathbf{X}$.}
\item{\makebox[2cm]{$\mathrm{diag}(\mathbf{x})$\hfill} Diagonal matrix with elements $\mathbf{x}$.}
\item{\makebox[2cm]{$\mathrm{span}(\mathbf{X})$\hfill} columnspace of matrix $\mathbf{X}$.}
\item{\makebox[2cm]{$\mathbf{I}$\hfill} Identity matrix.}
\item{\makebox[2cm]{$\mathbf{1}$\hfill} Ones matrix.}
\item{\makebox[2cm]{$\mathbf{0}$\hfill} Zeros matrix.}
\item{\makebox[2cm]{$\mathbf{x}^{y}$\hfill} Element-wise exponentiation.}
\item{\makebox[2cm]{$\mathcal{U}(a,b)$\hfill} Uniform distribution from $a$ to $b$.}
\end{itemize}

\section{Background: DOA Estimation for ULAs}\label{sec:overview}


\subsection{Model Framework}

Consider an array of $M$ sensors receiving uncorrelated signals from $K$ narrowband sources located in the far field of the array. Signals from each source arrive from angles $\boldsymbol{\theta} = [\theta_1\: \dots\:\theta_K]^\mathsf{T}$, which are given in radians. The sources are assumed to be in the same plane as the array, and the sensors are positioned at points on a line given by $\mathbf{r} = [r_1\:\dots\:r_M]^\mathsf{T}$, where each value $r_i$ denotes the distance of sensor $i$ from an arbitrary origin point in units of half-wavelengths. By using half-wavelengths as the primary unit of distance, many of the equations surrounding DOA estimation are simplified. If each sensor records $L$ snapshots of data, the measured signal can be modeled as
\begin{equation}\label{eq:model}
\begin{split}
\mathbf{Y} &= \mathbf{Z}+\mathbf{N}, \\
\mathbf{Z} &= \mathbf{A}_s(\mathbf{r},\boldsymbol{\theta})\mathbf{X},
\end{split}
\end{equation}
where $\mathbf{Y}\in \mathbb{C}^{M\times L}$ are the recorded measurements, $\mathbf{X} = [\mathbf{x}_1\:\dots\:\,\mathbf{x}_K]^{\mathsf{T}}\in\mathbb{C}^{K\times L}$ contains the signals from each of $K$ sources, Gaussian uncorrelated measurement noise is contained in $\mathbf{N}\in\mathbb{C}^{M\times L}$, and 
\begin{gather}\label{eq:ASM}
\mathbf{A}_s(\mathbf{r},\boldsymbol{\theta}) = [\mathbf{a}(\mathbf{r},\theta_1)\:\dots\:\,\mathbf{a}(\mathbf{r},\theta_K)], \\[7pt] \label{eq:manifold}
\mathbf{a}(\mathbf{r}, \theta) =[e^{-j{\pi} {r}_1\sin\theta}\:\,\dots\:\, e^{-j\pi{r}_M\sin\theta}]^\mathsf{T},
\end{gather}
is the array steering matrix whose columns model the phase pattern across the array from a signal arriving at angle $\theta$. The column vectors are known as \textit{array steering vectors} and are defined over $\theta \in [-\frac{\pi}{2},\frac{\pi}{2})$. The goal is to recover $\boldsymbol{\theta}$ given only knowledge of the sensor positions $\mathbf{r}$ and measurements $\mathbf{Y}$. 


\subsection{Vandermonde and Toeplitz Matrices}

A Vandermonde matrix $\mathbb{C}^{M\times K}$ is defined as 
%
%
\begin{equation}\label{eq:vand}
\begin{split}
\mathbf{V}(\boldsymbol{z}) &= [\boldsymbol{z}^0\: \boldsymbol{z}^1\:\dots\:\boldsymbol{z}^{(M-1)}]^{\mathsf{T}},\\
&= [\mathbf{v}(z_1)\: \dots \: \mathbf{v}(z_K)],
\end{split}
\end{equation}
where $\boldsymbol{z} = [z_1\:\dots\:z_K]^{\mathsf{T}}$ fully parameterizes the Vandermonde matrix \cite[p.~409]{moon2000mathematical}. A single column of the Vandermonde matrix is defined as $\mathbf{v}(z) = [1\: z^1\: \dots\:z^{M-1}]^{\mathsf{T}}$. 

The array steering matrix of a ULA will have Vandermonde structure. Sensor positions of a ULA are described as 
\begin{equation}\label{eq:rULA}
\mathbf{r}_{_\mathrm{ULA}} = \alpha [0\:\, 1\:\, \dots\:\, M-1]^{\mathsf{T}} + \beta,
\end{equation}
where $\alpha$ is an arbitrary scaling parameter and $\beta$ is an arbitrary shifting parameter. In this case the $\boldsymbol{\theta}$ parameters of \eqref{eq:ASM} are related to the $\boldsymbol{z}$ parameters of \eqref{eq:vand} by 
 \begin{gather}\label{eq:relate}
 {z}_k = e^{-j\pi(\alpha+\beta)\sin({\theta}_k)}, \\ \label{eq:relate_back} 
 {\theta}_k = -\sin^{-1}\Big(\frac{ \angle {z}_k}{(\alpha+\beta)\pi}\Big).
 \end{gather}
The values of $\mathbf{r}_{_\mathrm{ULA}}$ manifest as scaled and shifted integers. The classical choice of parameters has $(\alpha,\beta) = (1,0)$, and corresponds to a ULA with half-wavelength sensor spacing.

It is well known that any Toeplitz matrix, $\boldsymbol{\mathcal{T}}$, can be decomposed into Vandermonde components,
\begin{equation}\label{eq:HD}
\boldsymbol{\mathcal{T}} = \mathbf{V}(\boldsymbol{z})\mathbf{D}\mathbf{V}(\boldsymbol{z})^{\mathsf{H}},
\end{equation}
where $\mathbf{D} \in \mathbb{R}^{K\times K}$ is diagonal with positive entries, the $\boldsymbol{z}$ parameters have unit magnitude, and $\boldsymbol{\mathcal{T}}$ is a Hermitian symmetric Toeplitz matrix defined as 
\begin{equation}\label{eq:setD}
\boldsymbol{\mathcal{T}}(\mathbf{u}) =
\begin{pmatrix}
u_1 & u_2^* & \cdots & u_{M}^*\\
u_2 & u_1 &  \cdots & u_{M-1}^*\\
\vdots & \vdots & \ddots & \vdots\\
u_{M} & u_{M-1} & \cdots  & u_1\\
\end{pmatrix}.
\end{equation}
The first column, $\mathbf{u}$, of $\boldsymbol{\mathcal{T}}$ fully parameterizes the matrix. The decomposition of \eqref{eq:HD} is unique when the Toeplitz matrix is not full rank \cite{pisarenko1973retrieval}. The set of Toeplitz matrices will be denoted by the calligraphic letter $\boldsymbol{\mathcal{T}}$. 

The decomposition \eqref{eq:HD} will here be referred to as Vandermonde decomposition. 

\subsection{Root-MUSIC and Vandermonde Decoposition}\label{sec:root-MUSIC}

Root-MUSIC  is a subspace based DOA algorithm \cite{barabell1983improving, rao1989performance}. The algorithm consists of forming a polynomial from the sample covariance matrix of ULA measurements. The complex roots of the polynomial, $\boldsymbol{z}$, are used to estimate the DOAs through \eqref{eq:relate_back}. Without noise, the sample covariance matrix is Toeplitz and the polynomial roots are the $\boldsymbol{z}$ parameters of its Vandermonde decomposition \eqref{eq:HD}. Root-MUSIC is valid only for measurements taken at a ULA. 

\begin{figure}[t]
\begin{minipage}[t]{\linewidth}
\centering
\centerline{\includegraphics[width=\textwidth]{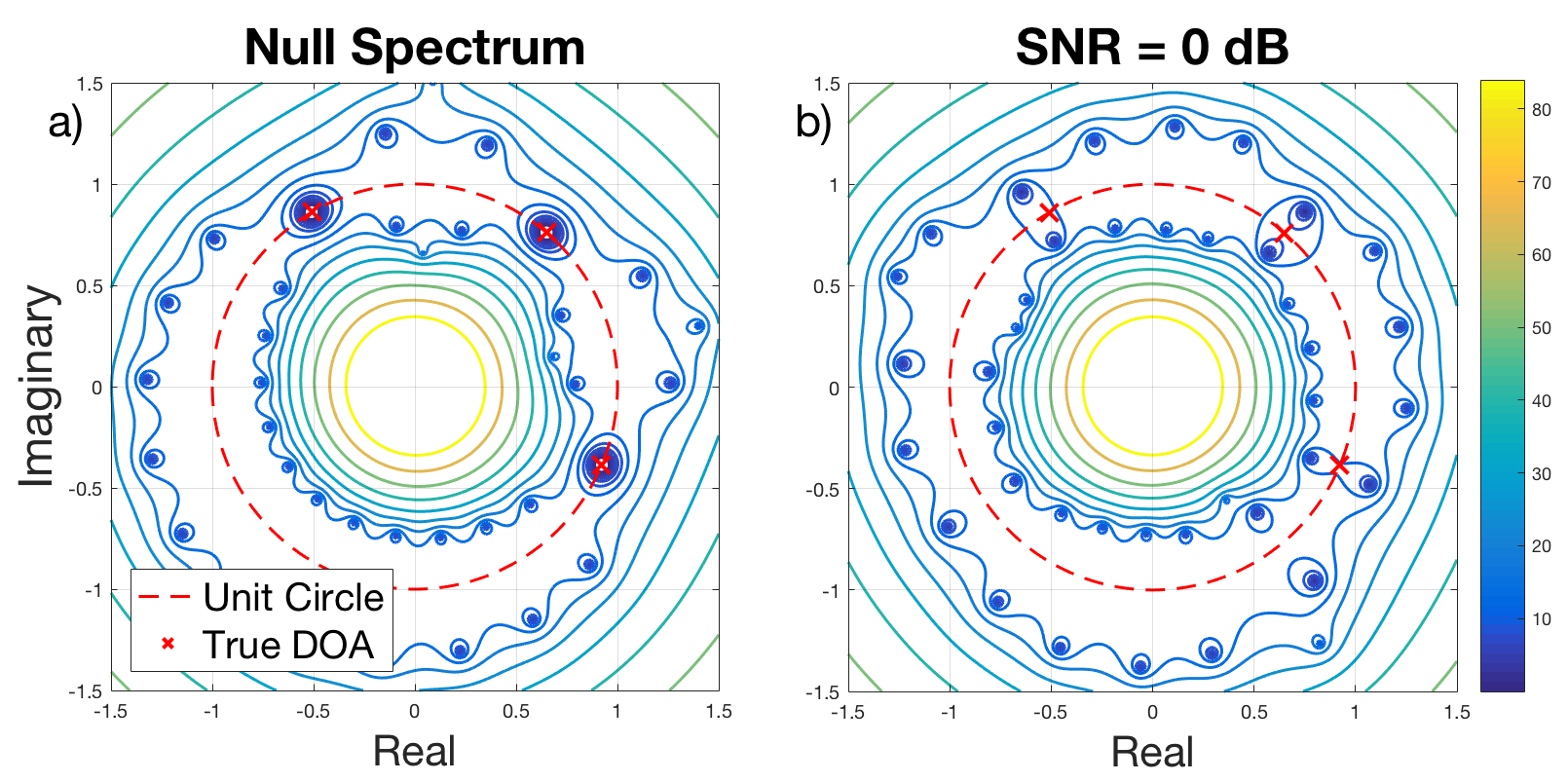}}
\caption{Null spectrum, \eqref{eq:polynomial}, for half-wavelength spaced ULA measurements generated according to \eqref{eq:model} using $M = 21$, $K = 3$, $L = 10$. DOAs located at $\boldsymbol{\theta} = [-7.2,15.9, 42.1]^{\circ}$ (red x).  a) Null spectrum contour of noiseless measurements. Red dashed line marks the complex unit circle. b) Same as a) with added noise such that $\| \mathbf{Z}\|_F = \| \mathbf{N}\|_F$, see \eqref{eq:model}.}
\label{fig:NSF1}
\end{minipage}
\end{figure}

The sample covariance matrix, $\mathbf{R}_{yy}$, is 
\begin{equation}\label{eq:Ryy1}
\begin{aligned}
\mathbf{R}_{yy} &= \frac{1}{L}\mathbf{YY}^{\mathsf{H}}, \quad \mathbb{E}[\mathbf{R}_{yy}] = \big(\mathbf{A}_s\boldsymbol{\Sigma}_{X}\mathbf{A}_s^{\mathsf{H}}+ \mathbf{\Sigma}_N\big), \\[7pt]
\mathbf{\Sigma}_X &=\frac{1}{L}\mathbb{E}\Big[\mathbf{XX}^{\mathsf{H}}\Big], \quad \mathbf{\Sigma}_N = \frac{1}{L}\mathbb{E}\Big[\mathbf{N}\mathbf{N}^{\mathsf{H}}\Big] = \sigma_N^2\mathbf{I},
\end{aligned}
\end{equation}
where $\mathbf{\Sigma}_N$ is the expected noise covariance matrix, which becomes $\sigma_N^2\mathbf{I}$ for uncorrelated noise with variance $\sigma_N^2$. 

If the measurements are noiseless, then $\mathrm{span}(\mathbf{R}_{yy}) = \mathrm{span}(\mathbf{A}_s)$. In this case, the columnspace and left nullspace of $\mathbf{R}_{yy}$ are known as the \textit{signal and noise subspaces} respectively. When noise is present in the measurements an estimated basis to both spaces is found from the eigen-decomposition of $\mathbf{R}_{yy}$,
\begin{equation}\label{eq:EVD}
\mathbf{R}_{yy} = \mathbf{U}_S\boldsymbol{\Lambda}_S\mathbf{U}_S^{\mathsf{H}} + \mathbf{U}_N\boldsymbol{\Lambda}_N\mathbf{U}_N^{\mathsf{H}},
\end{equation}
where $\boldsymbol{\Lambda}_S\in \mathbb{R}^{K\times K}$ is diagonal with the $K$ largest eigenvalues of $\mathbf{R}_{yy}$, $\mathbf{U}_S\in\mathbb{C}^{M\times K}$ is a matrix whose columns are the corresponding eigenvectors, and $\boldsymbol{\Lambda}_N\in\mathbb{R}^{(M-K) \times (M-K)}$ and $\mathbf{U}_N\in\mathbb{C}^{M\times(M-K)}$ are matrices containing the remaining noise eigenvalues and eigenvectors. The matrices $\mathbf{U}_S$ and $\mathbf{U}_N$ are estimated bases to the signal and noise subspaces respectively. It must be noted that the signal and noise subspaces can only be estimated when $L\geq K$, because at least $K$ snapshots are required to build a rank $K$ approximation of $\mathbf{R}_{yy}$.

For noise free measurements, $\mathbf{U}_N$ will be composed of eigenvectors whose corresponding eigenvalues are 0. In this case $\mathbf{U}_N$ exactly spans the left nullspace of $\mathbf{A}_s$ and
\begin{equation}\label{eq:purp}
\mathbf{U}_N \perp \mathbf{A}_s \implies \mathbf{U}_N^{\mathsf{H}}\mathbf{A}_s = \mathbf{0},
\end{equation}
When noise is present, the eigenvectors composing $\mathbf{U}_N$ will have non-zero corresponding eigenvalues and $\mathbf{U}_N$ will be an approximation of the noise subspace. 

Because the array is assumed uniform and linear, any array steering vector can be substituted with a column from a Vandermonde matrix. From \eqref{eq:purp}, the \textit{null spectrum} is formed \cite[p.1159]{van2004optimum}
\begin{equation}\label{eq:nullspec}
\begin{aligned}
D({z}) = \|\mathbf{U}_N^{\mathsf{H}}\mathbf{a}(\theta)\|_2^2 &= \mathbf{a}({\theta})^{\mathsf{H}}\mathbf{U}_N\mathbf{U}_N^{\mathsf{H}}\mathbf{a}({\theta}), \\
&= \mathbf{v}(\frac{1}{z})^{\mathsf{T}}\mathbf{U}_N\mathbf{U}_N^{\mathsf{H}}\mathbf{v}({z}),
\end{aligned}
\end{equation}
for $|z| = 1$. If we define the matrix 
\begin{equation}\label{eq:G}
\mathbf{G} = \mathbf{U}_N\mathbf{U}_N^{\mathsf{H}},
\end{equation}
for $\mathbf{G}\in\mathbb{C}^{M\times M}$, then the null spectrum can be expanded into a polynomial in $z$,
\begin{equation}\label{eq:polynomial}
\begin{aligned} 
&d_i = \sum_{m_1,m_2}^M \mathbf{G}_{m_1,m_2}, \quad (m_1-m_2) = i, \\[5pt]
D({z}) = & \:d_{-(M-1)}{z}^{-(M-1)}+\dots+d_{M-1}{z}^{M-1}, \\[7pt]
\end{aligned}
\end{equation}
where $d_i$ are the sums along the diagonals of $\mathbf{G}$. Some properties of $D(z),\: z\in\mathbb{C},$ are as follows: 
\vspace{1.5mm}
\begin{enumerate}[i.]
\item\label{enum:1} $d_i = d_{-i}^*$, because $\mathbf{G}$ is Hermitian by construction. \vspace{.75mm}
\item\label{enum:2} As a result of \eqref{eq:purp}, $K$ root pairs appear near the unit circle (noise present), or on the unit circle (noise free) \cite{rao1989performance}.  \vspace{.75mm}
\item\label{enum:3} If one root exists at $\tilde{z}$, another root will exist at $\frac{1}{\tilde{z}^*}$ because $D(z) = D(\frac{1}{z^*})$. \vspace{.75mm}
\end{enumerate}
\vspace{1.5mm}

For root-MUSIC, the $K$ roots inside the unit circle with largest magnitude, $|z|$, are taken as DOA estimates using the mapping of \eqref{eq:relate_back} \cite{van2004optimum}. These roots produce highly reliable estimates of the DOAs. When noise is present in the measurements, the accuracy of the DOA estimates is related to how well the true noise subspace was approximated by $\mathbf{U}_N$.

Note that $D(z)$ was derived only for $z$ on the unit circle, but its roots located off the unit circle are used as DOA estimates. There is no physically meaningful reason behind this decision. Rather, it should be viewed as a useful mathematical trick. The polynomial expansion of $D(z)$ is especially useful because its roots can be calculated efficiently. 

Figure \ref{fig:NSF1} depicts an example null spectrum from half-wavelength spaced ULA measurements, which highlights points \ref{enum:2} and \ref{enum:3}. Even when the measurements are severely corrupted by noise, the null spectrum can be used to obtain good estimates of the DOA locations. 

When $\mathbf{R}_{yy}$ is noise free, it is rank $K$ and takes Toeplitz structure. Additionally, $K$ (double) roots appear on the unit circle of its null spectrum. Those roots are the elements, $\boldsymbol{z}$, of the Vandermonde decomposition of $\mathbf{R}_{yy}$, see \eqref{eq:HD}. Thus the procedure of root-MUSIC finds the $\boldsymbol{z}$ parameters of a Vandermonde decomposition. Because root-MUSIC works for covariance matrices corrupted by noise, it can be viewed as a method for approximating the $\boldsymbol{z}$ parameters of a `noisy' Toeplitz matrix.



\subsection{Gridless DOA for Uniform Linear Arrays}

In gridless DOA estimation, the DOAs composing $\mathbf{Z}$, \eqref{eq:model}, are found by minimizing the atomic $\ell_0$ norm of atoms defined by the manifold of the array steering matrix \cite{YANG2018509}.

The noiseless signal contained in $\mathbf{Z}$ can be re-written as
\begin{equation}
\mathbf{Z} = \sum_{k=1}^K\mathbf{a}(\mathbf{r}, \theta_k)\mathbf{x}_k^{\mathsf{H}} = \sum_{k=1}^K\mathit{c}_k\mathbf{a}(\mathbf{r},\theta_k)\mathbf{b}_k^{\mathsf{H}}
\end{equation}
where $\mathit{c}_k = \| \mathbf{x}\|_2 > 0$, $\mathbf{b}_k = \mathit{c}_k^{-1}\mathbf{x}_k$, thus $\| \mathbf{b}_k \|_2 = 1$. We define the atomic set as
\begin{equation}\label{eq:atomic_set}
\mathcal{A} = \{ \mathbf{a}(\mathbf{r},\theta_k,\mathbf{b}_k) = \mathbf{a}(\mathbf{r},\theta_k)\mathbf{b}_k^{\mathsf{H}}\},
\end{equation}
which can be thought of as the set of rank 1 matrices of constrained norm that can be constructed from the manifold $\mathbf{a}(\mathbf{r}, \theta)$ over all values of $\theta$. Expressing $\mathbf{Z}$ as a linear combination of $K$ atoms in $\mathcal{A}$ brings us to the definition of the atomic $\ell_0$ norm formulation of $\mathbf{Z}$,
\begin{equation}\label{eq:ANM}
\begin{aligned}
\lVert\mathbf Z\rVert_{\mathcal{A},0} 
&=\inf_{\mathit{c}_k,\theta_k,\mathbf{b}_k} \Bigg\{ K:\: \mathbf{Z} = \sum_{k=1}^{K} \mathit{c}_k\mathbf{a}(\mathbf{r},\theta_k)\mathbf{b}_k^{\mathsf{H}}\Bigg\}.
\end{aligned}
\end{equation}

Gridless DOA is concerned with finding a solution to \eqref{eq:ANM}. Towards this goal, consider the matrix
\begin{equation}\label{eq:S}
\mathbf{S} =  \sum_{k=1}^K c_k^2\begin{bmatrix}
\mathbf{a}(\mathbf{r},\theta_k) \\
\mathbf{b}_k
\end{bmatrix} \begin{bmatrix}
\mathbf{a}(\mathbf{r},\theta_k) \\
\mathbf{b}_k
\end{bmatrix}^\mathsf{H} = \begin{bmatrix} 
\mathbf{T} & \mathbf{Z} \\
\:\mathbf{Z}^{\mathsf{H}} & \mathbf{Q} 
\end{bmatrix},
\end{equation}
where 
%
\begin{align}\label{eq:T}
\mathbf{T} &= \mathbf{A}_s(\mathbf{r},\boldsymbol{\theta})\mathbf{D}\mathbf{A}_s(\mathbf{r},\boldsymbol{\theta})^{\mathsf{H}}, \\ 
\mathbf{Q} &= \mathbf{X}^{\mathsf{H}}\mathbf{X},
\end{align}
%
and $\mathbf{D}\in\mathbb{R}^{K\times K}$ is diagonal with elements $c_k^2$. By definition $\mathbf{S}$ is a positive semi-definite (PSD) matrix. 

In the ULA case the array steering matrix is Vandermonde and $\mathbf{T} \in\boldsymbol{\mathcal{T}}$. Theorem 6.2 of \cite{YANG2018509} tells us that the atomic $\ell_0$ norm of \eqref{eq:ANM} in the ULA case will be the optimal solution of the following rank constrained optimization problem,
\begin{equation}\label{eq:gridless2}
\begin{aligned}
& \underset{\mathbf{T}\in\boldsymbol{\mathcal{T}},\:\mathbf{Q}}{\text{minimize}}
& & \mathrm{rank}\big(\mathbf{{T}}\big),  \quad \text{subject to} \quad \mathbf{S}\succeq 0.
\end{aligned}
\end{equation}
Once the optimal $\mathbf{{T}}$ is found, the DOAs $\theta_k$ for $k = 1,\dots,K$ can be recovered through Vandermonde decomposition (or root-MUSIC) of $\mathbf{T}$. Thus \eqref{eq:gridless2} can be viewed as a means of estimating the full rank covariance matrix of a single measurement snapshot. This is possible because the covariance matrix has low rank Toeplitz structure.

State-of-the-art optimization solvers are only suitable for convex problems. In practice, the non-convex optimization problem of \eqref{eq:gridless2} is substituted for its convex relaxation \cite{YANG2018509},
\begin{equation}\label{eq:gridless_cvx}
\begin{aligned}
& \underset{\mathbf{T}\in\boldsymbol{\mathcal{T}},\:\mathbf{Q}}{\text{minimize}}
& & \mathrm{Tr}\big(\mathbf{{T}}\big) + \mathrm{Tr}\big(\mathbf{{Q}}\big),  \quad \text{subject to} \quad \mathbf{S}\succeq 0.
\end{aligned}
\end{equation}
When the matrix rank is substituted for the matrix trace the optimization of \eqref{eq:gridless2} is in the form of a semi-definite program (SDP), to which there are many available solvers \cite{grant2008cvx}. The derivation of the popular alternating directions method of multipliers (ADMM) algorithm \cite{boyd2004convex} to solve \eqref{eq:gridless_cvx} is provided in App. \ref{sec:Appendix}.



\section{Extension to Non-Uniform Array Geometries}\label{sec:EVD}

We introduce a generalization of the Vandermonde matrix, which we call the \textit{irregular Vandermonde matrix}. In the same way a Toeplitz matrix is constructed from Vandermonde components, we define an \textit{irregular Toeplitz matrix} constructed from irregular Vandermonde components. It is then shown that irregular Toeplitz matrices can be decomposed back to their irregular Vandermonde components. 

We deem this decomposition the \textit{irregular Vandermonde decomposition} (IVD) because it can be interpreted as the Vandermonde decomposition of an irregularly sampled signal (i.e. the positions of the sensors act as sample locations of a spectrally sparse signal whose frequencies are related to the DOAs). Furthermore, the array steering matrix of a NUA has irregular Vandermonde structure. We propose an `irregular' root-MUSIC algorithm, which is related to the IVD in the same way root-MUSIC is related to the Vandermonde decomposition.

The IVD is derived by carrying out the steps of the Vandermonde decomposition on the irregular Vandermonde matrix. Under this framework, the null spectrum no longer has polynomial structure and cannot be easily rooted. Regardless, the roots of interest are the $2K$ root pairs which lie on or near the unit circle. A simple method to recover the relevant information from these roots is presented, which does not resort to computationally expensive numerical methods.

The key difference between the Vandermonde decomposition and the IVD is an additional vector parameter involved in the IVD which specifies the sensor positions. For any vector of sensor positions there is a set of matrices akin to the Toeplitz set containing all matrices that can be decomposed exactly by the IVD. This set is used in Sec. \ref{sec:contribution} to extend gridless DOA to NUA measurements. 

\subsection{Irregular Vandermonde and Toeplitz Matrices}

Consider an irregular Vandermonde matrix $\mathbb{C}^{M \times K}$ defined as \cite{demmel2005accurate},
%
%
\begin{equation}
\begin{split}
\mathbf{W}(\boldsymbol{\gamma},\boldsymbol{z}) &= [\boldsymbol{z}^{\gamma_1}\dots\:\boldsymbol{z}^{\gamma_{M}}]^{\mathsf{T}}, \\[5pt]
&= [\mathbf{w}(\boldsymbol{\gamma},z_1)\:\dots\: \mathbf{w}(\boldsymbol{\gamma},z_K)],
\end{split}
\end{equation}
where $\gamma_i$ is the $i$th element of a vector $\boldsymbol{\gamma}\in\mathbb{R}^M$ and $\boldsymbol{z}\in\mathbb{C}^K$, and $\mathbf{w}(\boldsymbol{\gamma},z) = [z^{\gamma_1}\:\dots\:z^{\gamma_M}]^{\mathsf{T}}$. 

In the context of DOA estimation, $\mathbf{A}_s(\mathbf{r},\boldsymbol{\theta}) = \mathbf{W}(\boldsymbol{\gamma},\boldsymbol{z}) $, using the mapping
%
\begin{equation}\label{eq:zdef}
 \boldsymbol{\gamma} = \mathbf{r}, \quad {z}_k = e^{-j\pi\sin{\theta}_k},\quad \theta_k = -\sin^{-1}(\frac{\angle z_k}{\pi}). 
\end{equation}
%
Notice that this is a generalization of the mapping presented in (\ref{eq:rULA}--\ref{eq:relate_back}).

Following \eqref{eq:HD}, we  construct an irregular Toeplitz matrix from irregular Vandermonde matrices in the same way a Toeplitz matrix is constructed from Vandermonde matrices.
Define $\boldsymbol{\mathcal{T}}_{\subVecgamma}$ as the irregular Toeplitz set for parameter vector $\boldsymbol{\gamma}$,
\begin{equation}\label{eq:setA}
\boldsymbol{\mathcal{T}}_{\subVecgamma} = \{ \mathbf{T} : \mathbf{T} = \mathbf{W}(\boldsymbol{\gamma},\boldsymbol{z})\mathbf{D} \mathbf{W}(\boldsymbol{\gamma},\boldsymbol{z})^{\mathsf{H}}, |\boldsymbol{z}| = \boldsymbol{1} \},
\end{equation}
where $\mathbf{D}\in\mathbb{R}^{K\times K}$ is a diagonal matrix with elements $c_k^2$. Each unique parameter vector, $\boldsymbol{\gamma}$, is associated with an irregular Toeplitz set, $\boldsymbol{\mathcal{T}}_{\subVecgamma}$. The Toeplitz set is reached by any parameter vector corresponding to a ULA, $\boldsymbol{\gamma} = \alpha[0,\dots,M-1]^{\mathsf{T}} + \beta$.

The expected sample covariance matrix for NUA measurements with sensor positions $\mathbf{r} = \boldsymbol{\gamma}$ is a diagonally loaded member of $\boldsymbol{\mathcal{T}}_{\subVecgamma}$,
\begin{equation}\label{eq:covar}
\begin{aligned}
\mathbb{E}[\mathbf{R}_{yy}] &= \mathbf{A}_s(\mathbf{r},\boldsymbol{\theta})\mathbf{\Sigma}_X\mathbf{A}_s(\mathbf{r},\boldsymbol{\theta})^{\mathsf{H}} + \mathbf{\Sigma}_{N} \\
&= \mathbf{W}(\boldsymbol{\gamma},\boldsymbol{z})\mathbf{\Sigma}_{X} \mathbf{W}(\boldsymbol{\gamma},\boldsymbol{z})^{\mathsf{H}} + \sigma_N^2\mathbf{I}.
\end{aligned}
\end{equation}
for $\mathbf{\Sigma}_{X}$ and $\mathbf{\Sigma}_{N}$ in \eqref{eq:Ryy1}. When the measurements are noiseless $\mathbf{\Sigma}_N = \mathbf{0}$ and $\mathbf{R}_{yy}\in \boldsymbol{\mathcal{T}}_{\subVecgamma}$. 

\begin{figure}[t]
\begin{minipage}[t]{\linewidth}
\centering
\centerline{\includegraphics[width=\textwidth]{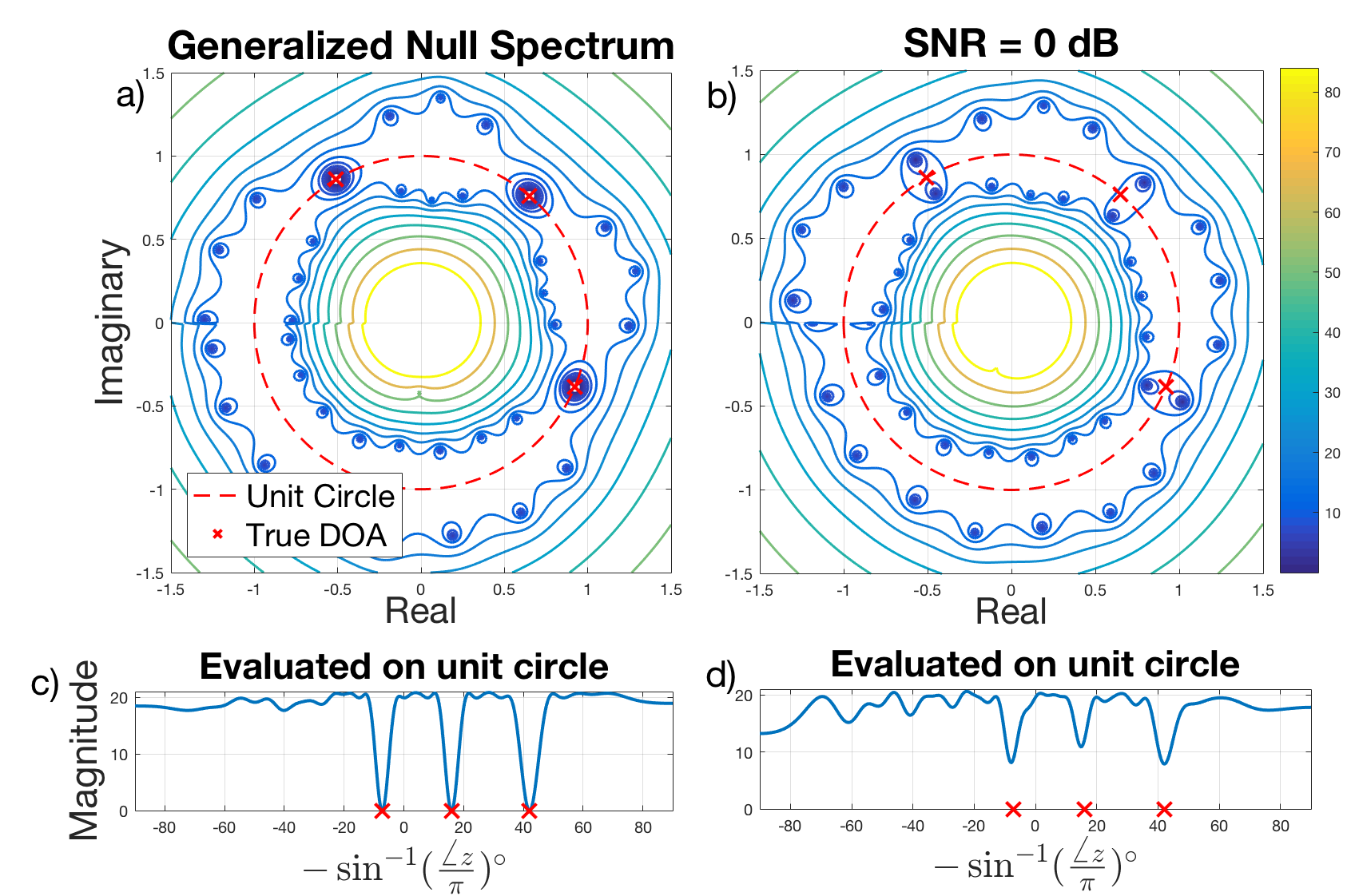}}
\caption{Irregular null spectrum, \eqref{eq:poly}, for NUA measurements generated according to \eqref{eq:model} using $M = 21$, $K = 3$, $L = 10$. DOAs located at $\boldsymbol{\theta} = [-7.24,15.96, 42.07]^{\circ}$ (red x).  a) Irregular null spectrum contour of noiseless measurements. Red dashed line marks the complex unit circle. b) Same as a) with added noise such that $\|\mathbf{Z}\|_F = \| \mathbf{N}\|_F$, see \eqref{eq:model}. c) Evaluation of a) on unit circle. d) Evaluation of b) on unit circle.}
\label{fig:NSF2}
\end{minipage}
\end{figure}

\subsection{Irregular Root-MUSIC and Vandermonde Decomposition} \label{sec:roots}

Consider the sample covariance matrix, $\mathbf{T}\in \mathbb{C}^{M\times M}$, from a NUA constructed as in \eqref{eq:covar}. Following \eqref{eq:nullspec}, the null spectrum of $\mathbf{T}$ is 
\begin{equation}\label{eq:nullspec2}
\begin{aligned}
\tilde{D}({z}) = \|\mathbf{U}_N^{\mathsf{H}}\mathbf{a}(\mathbf{r},\theta)\|_2^2 &= \mathbf{a}(\mathbf{r},\theta)^{\mathsf{H}}\mathbf{U}_N\mathbf{U}_N^{\mathsf{H}}\mathbf{a}(\mathbf{r},\theta) \\
&= \mathbf{w}(\boldsymbol{\gamma},\frac{1}{z})^{\mathsf{T}}\mathbf{G}\mathbf{w}(\boldsymbol{\gamma},{z}),
\end{aligned}
\end{equation}
for $\mathbf{r} = \boldsymbol{\gamma}$ and $\mathbf{G}$ in \eqref{eq:G}. Because \eqref{eq:nullspec2} is constructed from irregular Vandermonde matrices, we refer to it as the \textit{irregular null spectrum}. The expansion of $\tilde{D}(z)$ is
\begin{equation}\label{eq:poly}
\tilde{D}({z}) = \sum_{m= 1}^M \sum_{n=1}^M g_{m,n}z^{\gamma_m-\gamma_n},
\end{equation}
where $g_{m,n}$ is element $(m,n)$ of $\mathbf{G}$. We treat the domain of $\tilde{D}(z)$ as though it were the set of complex numbers, despite deriving $\tilde{D}(z)$ only for $z$ on the unit circle.

Figure \ref{fig:NSF2} depicts an example irregular null spectrum, $\tilde{D}(z)$, from NUA measurements. The behavior of the irregular null spectrum is similar to that of the null spectrum from a half-wavelength spaced ULA in Fig. \ref{fig:NSF1}. Note that the irregular null spectrum has a discontinuity at $\angle z = \pi$, because for nearly all array geometries $\tilde{D}(z)\rvert_{\angle z = \pi_+} \neq \tilde{D}(z)\rvert_{\angle z = \pi_-}$. 

To understand why this is true, consider that $\angle z = \pi_{\pm}$ corresponds to DOAs $\theta=\pm\frac{\pi}{2}$, \eqref{eq:zdef}. The discontinuity is due to the inequality of the NUA array manifold at the extreme values of $\theta$. The half-wavelength spaced ULA is one of several exceptional array geometries which produce the same array pattern across the array for signals arriving at $\pm 90^{\circ}$,
\begin{equation}
e^{-j\pi (r+\beta) \sin(\frac{\pi}{2})} = e^{-j\pi (r+\beta) \sin(-\frac{\pi}{2}) }, \quad r \in \mathbb{Z}.
\end{equation}

%
%

Properties \ref{enum:1}-\ref{enum:3} from Sec. \ref{sec:root-MUSIC} hold for the irregular null spectrum, (the equivalent property \ref{enum:1} for the irregular null spectrum is $g_{m,n} = g_{n,m}^*$). Unlike \eqref{eq:polynomial}, the irregular null spectrum no longer expands to a polynomial that can be easily rooted. Instead, the expansion of \eqref{eq:poly} is a non-linear equation for which there is no known closed form solution to obtain its roots.

We exploit the following two facts about $\tilde{D}(z)$:
\begin{enumerate}[1.]
\item\label{enum2:1} The roots of interest are those that appear on or near the unit circle, as seen by \eqref{eq:purp}. \vspace{.75mm}
\item\label{enum2:2} Only the phase angle of the roots of interest are used to generate DOA estimates. \vspace{.75mm}
\end{enumerate}
Thus the local minimums of $\tilde{D}(z)$ evaluated on the unit circle give DOA estimates with similar accuracy as those given by the actual roots. $\tilde{D}(z)$ evaluated on the unit circle is the inverse of the MUSIC spectrum \cite{schmidt1986multiple}. 

The arguments, $\boldsymbol{z}$, producing the $K$ smallest local minima of $\tilde{D}(z)$ are taken as the DOA estimates using \eqref{eq:zdef}. Because each root of $\tilde{D}(z)$ has a root pair on the same radial line (i.e. $\angle \tilde{z} = \angle \frac{1}{\tilde{z}^*}$), a saddle point is expected to lie on or near this radial line. These saddle points manifest as local minima of $\tilde{D}(z)$ evaluated over $|z| = 1$, and provide good estimates of the DOAs. An example is provided in Fig. \ref{fig:NSF2}.

When $\mathbf{T}\in \boldsymbol{\mathcal{T}}_{\subVecgamma}$, all roots of interest are guaranteed to appear on the unit circle because $\mathbf{G}\perp \mathbf{W}(\boldsymbol{\gamma},\boldsymbol{z})$. Evaluating $\tilde{D}(z)$ on the unit circle yields all root locations, $\boldsymbol{z}$, which perfectly reconstructs $\mathbf{W}(\boldsymbol{\gamma},\boldsymbol{z})$ from \eqref{eq:setA}. The signal powers, $\mathbf{c} = [c_1^2\:\dots\:c_K^2]^{\mathsf{T}}$ are found from,
\begin{equation}\label{eq:lininv}
\begin{aligned}
\mathbf{c} = \mathrm{diag}&\Big(\mathbf{W}^{\dagger}\mathbf{T}(\mathbf{W}^{{\dagger}})^{\mathsf{H}}\Big), \\[6pt]
\mathbf{W}^{\dagger} = (&\mathbf{W}^{\mathsf{H}}\mathbf{W})^{-1}\mathbf{W}^{\mathsf{H}},
\end{aligned}
\end{equation}
where $\mathbf{D} = \mathrm{diag}(\mathbf{c})$, and $\mathbf{W}(\boldsymbol{\gamma},\boldsymbol{z})$ has been shortened to $\mathbf{W}$. By this procedure it is possible to decompose an irregular Toeplitz matrix to irregular Vandermonde components. Pseudocode for the IVD algorithm is given in Sec. \ref{sec:algs}.

\subsection{Irregular Toeplitz Structure}

This section provides insight into the specific structure shared by all members of $\boldsymbol{\mathcal{T}}_{\subVecgamma}$, similar to Toeplitz matrices. Consider a noiseless sample covariance matrix, this time constructed as an irregular Toeplitz matrix,
\begin{equation}
\mathbf{R}_{zz} = \mathbf{A}_s(\mathbf{r},\boldsymbol{\theta})\mathbf{\Sigma}_X\mathbf{A}_s(\mathbf{r},\boldsymbol{\theta})^{\mathsf{H}} = \mathbf{W}(\boldsymbol{\gamma},\boldsymbol{z})\mathbf{D}\mathbf{W}(\boldsymbol{\gamma},\boldsymbol{z})^{\mathsf{H}},
\end{equation}
where $\mathbf{\Sigma}_X = \mathbf{D} = \mathrm{diag}(\mathbf{c})$, and $(\boldsymbol{\gamma},{z})$ is related to $(\mathbf{r},{\theta})$ by \eqref{eq:zdef}. Each element of $\mathbf{R}_{zz}$ can be written as,
\begin{equation}
(\mathbf{R}_{zz})_{m,n} = \sum_{k=1}^{K}c_k^2 z_k^{(\gamma_m-\gamma_n)}.
\end{equation}
In the ULA case, $\boldsymbol{\gamma} = \alpha[0,1,\dots,(M-1)]^{\mathsf{T}}+\beta$, and it can be seen that $\gamma_m-\gamma_n$ will take the same value across each diagonal, producing the Toeplitz structure. In the NUA case element $(m,n)$ of $\mathbf{R}_{zz}$ is the sum of $K$ complex exponential functions sampled at element $(m,n)$ of the Euclidean distance matrix of $\boldsymbol{\gamma}$ \cite{dattorro2010convex}.

It can be shown that the irregular Toeplitz set for any $\boldsymbol{\gamma}$ is convex. Consider two matrices in $\boldsymbol{\mathcal{T}}_{\subVecgamma}$ composed from irregular Vandermonde matrices with parameters $\boldsymbol{z}_1$ and $\boldsymbol{z}_2$,
\begin{align}
\mathbf{T}_1 &= \mathbf{W}(\boldsymbol{\gamma},\boldsymbol{z}_1)\mathbf{D}_1\mathbf{W}(\boldsymbol{\gamma},\boldsymbol{z}_1)^{\mathsf{H}}, \\
\mathbf{T}_2 &= \mathbf{W}(\boldsymbol{\gamma},\boldsymbol{z}_2)\mathbf{D}_2\mathbf{W}(\boldsymbol{\gamma},\boldsymbol{z}_2)^{\mathsf{H}}.
\end{align}
Then their convex combination is
\begin{equation}\label{eq:convex}
\begin{split}
\lambda\mathbf{T}_1+&(1-\lambda)\mathbf{T}_2 = \\
& \mathbf{W}\Big(\boldsymbol{\gamma},[\boldsymbol{z}_1\:\boldsymbol{z}_2]\Big)\begin{bmatrix} \lambda\mathbf{D}_1&\boldsymbol{0}\\\boldsymbol{0}&(1-\lambda)\mathbf{D}_2 \end{bmatrix}\mathbf{W}\Big(\boldsymbol{\gamma},[\boldsymbol{z}_1\:\boldsymbol{z}_2]\Big)^{\mathsf{H}},
\end{split}
\end{equation}
for $0\leq\lambda\leq 1$. The right hand side of \eqref{eq:convex} is a member of $\boldsymbol{\mathcal{T}}_{\subVecgamma}$ composed of irregular Vandermonde matrices with parameters $[\boldsymbol{z}_1 \: \boldsymbol{z}_2]$.

\subsection{IVD and Irregular Root-MUSIC}\label{sec:algs}

The IVD presents a computationally efficient extension of root-MUSIC to NUA measurements. We present pseudocode for the algorithms here. 

The IVD takes as input the sensor positions, $\boldsymbol{\gamma}$, the dimension of the noise subspace, $K$, and the matrix to be decomposed, $\mathbf{T}$. The outputs are the harmonics, $\boldsymbol{z}$, and the diagonal matrix, $\mathbf{D}$. If $\mathbf{T}$ is a sample covariance matrix, then the harmonics are related to the DOAs by \eqref{eq:zdef}.
\begin{algorithm}[t]
\caption{: $(\boldsymbol{z},\mathbf{c}) = \mathtt{IVD}(\mathbf{T},\boldsymbol{\gamma},K)$}
\begin{algorithmic} 
\REQUIRE $\mathbf{T} \in \mathbb{C}^{M\times M}$, \: $\boldsymbol{\gamma}\in\mathbb{R}^{M}$, \: $K \in\mathbb{Z}$
\STATE $[ \mathbf{U},\boldsymbol{\Sigma},\mathbf{V} ] = \mathtt{svd}(\mathbf{T})$
\STATE $\mathbf{U}_N$= $\mathbf{U}(\: : \:,K+1:M)$ 
\STATE $\boldsymbol{z}$= $\mathtt{find}(\:\mathtt{argmin}(\:\tilde{D}(z)\:) , |z| = 1\:)$, \: \textrm{see}\:\eqref{eq:poly} and \eqref{eq:localmin} 
\STATE $\mathbf{c}$ = $\mathrm{diag}\Big(\mathbf{W}(\boldsymbol{\gamma},\boldsymbol{z})^{\dagger}\mathbf{T}(\mathbf{W}(\boldsymbol{\gamma},\boldsymbol{z})^{\dagger})^{\mathsf{H}}\Big)$
\end{algorithmic}\label{alger:IVD}
\end{algorithm}
In Algorithm \ref{alger:IVD}, $\mathtt{svd()}$ is the singular value decomposition of a matrix and $\mathtt{find}( \mathtt{argmin} (\tilde{D}(z)), |z| = 1 )$ outputs the values of $z$ producing the $K$ smallest local minima of $\tilde{D}(z)$ on the unit circle. For simulations in Sec. \ref{sec:Sim}, a gridded search over $10M$ evenly spaced points on the unit circle was used to find intervals containing minima, then the estimated minima locations were iteratively refined using a golden section search algorithm \cite{forsythe1977computer}.

The IVD allows for the extension of root-MUSIC to array measurements taken at a non-uniform array. We deem this \textit{irregular root-MUSIC}, which is presented in Algorithm \ref{alger:eRM}.
\begin{algorithm}[t]
\caption{: $\hat{\boldsymbol{\theta}} = \mathtt{IrregularRootMusic}(\mathbf{Y},\boldsymbol{\gamma},K)$}
\begin{algorithmic} 
\REQUIRE $\mathbf{Y} \in \mathbb{C}^{M\times L}$, \: $\boldsymbol{\gamma}\in\mathbb{R}^{M}$, \: $K \in\mathbb{Z}$,\: $L \geq K$
\STATE $\mathbf{T} = \frac{1}{L}\mathbf{Y}\mathbf{Y}^{\mathsf{H}} $
\STATE $(\boldsymbol{z},\mathbf{c}) = \mathtt{IVD}(\mathbf{T},\boldsymbol{\gamma},K)  $
\STATE $\hat{\boldsymbol{\theta}}  = -\mathtt{asin}(\mathtt{angle}({\boldsymbol{z}})/\pi)$
\end{algorithmic}\label{alger:eRM}
\end{algorithm}
Irregular root-MUSIC takes the measurements, $\mathbf{Y}$, as inputs and outputs the DOAs. The number of measurement snapshots must be at least as large as the number of DOAs present ($L\geq K$) or it will be impossible to accurately estimate the noise subspace and the resulting DOA estimates will be incorrect. The sensor positions, $\boldsymbol{\gamma}$, input to Algorithm \ref{alger:eRM} are in units of half-wavelengths.

\section{Gridless DOA for Non-Uniform Arrays} \label{sec:contribution}

The irregular Toeplitz set enables gridless DOA to be generalized to the NUA case. In this section we modify the optimization for gridless DOA to extend its use to measurements from NUAs, then propose an alternating projections (AP) based algorithm for solving said optimization. The AP solution is based on similar AP based algorithms that were recently examined in the context of CCS \cite{condat2015cadzow,cho2016fast,liu2017projected}.

\subsection{Extension to Non-Uniform Arrays}\label{sec:extension}

To extend gridless DOA to NUAs, the optimization of \eqref{eq:gridless2} is modified such that the optimal matrix belongs to $\boldsymbol{\mathcal{T}}_{\subVecgamma}$ for $\boldsymbol{\gamma} = \mathbf{r}$,
\begin{equation}\label{eq:gridlessarb1}
\begin{aligned}
& \underset{\mathbf{T}\in \boldsymbol{\mathcal{T}}_{\subVecgamma},\:\mathbf{Q}}{\text{minimize}}
& & \mathrm{rank}\big(\mathbf{T}\big), \quad \text{subject to}\:\: \mathbf{S}\succeq 0.\\
\end{aligned}
\end{equation}
%
The optimization of \eqref{eq:gridlessarb1} can be further simplified if we define the rank constrained irregular Toeplitz set as
\begin{equation}\label{eq:setA_rankK}
\boldsymbol{\mathcal{T}}_{\subVecgamma}^K = \{ \mathbf{T} : \mathbf{T} \in \boldsymbol{\mathcal{T}}_{\subVecgamma},\:\: \mathrm{rank}(\mathbf{T}) \leq K \}.
\end{equation}
This is the set of rank $K$ matrices that are members of $\boldsymbol{\mathcal{T}}_{\subVecgamma}$.
The NUA gridless DOA problem then simplifies to, 
\begin{equation}\label{eq:gridlessarb}
\begin{aligned}
& \underset{\mathbf{T}\in \boldsymbol{\mathcal{T}}_{\subVecgamma}^K,\:\mathbf{Q}}{\text{minimize}} \quad \|\mathbf{S}\|_F
 \quad \text{subject to}\:\: \mathbf{S}\succeq 0. \\
\end{aligned}
\end{equation}
Once the optimal $\mathbf{T}$ is known, the IVD (Algorithm \ref{alger:IVD}) can be applied to recover the DOAs and source powers.

\subsection{Important Projections NUA Gridless DOA}\label{sec:extVandset}
Before it is possible to formulate the optimization of \eqref{eq:gridlessarb} using AP, we define some important projections.

\subsubsection{Projection to the Toeplitz Set}
For ULA measurements, the projection onto $\boldsymbol{\mathcal{T}}_{\subVecgamma}^K$ can be replaced by a projection onto $\boldsymbol{\mathcal{T}}$. The projection of a matrix, $\mathbf{T}\in\mathbb{C}^{M\times M}$, to the Toeplitz set is \cite{eberle2003finding},
\begin{equation}
\begin{aligned}\label{eq:PToep}
P_{\boldsymbol{\mathcal{T}}}(\mathbf{T}) &= \boldsymbol{\mathcal{T}}(\mathbf{u}) \\[7pt]
u_i &= \frac{1}{2(M-i)}\sum_{j = 1}^{(M-i)}\mathbf{T}_{j,j+i-1} + \mathbf{T}_{j+i-1,j}^*.
\end{aligned}
\end{equation}
In words, the Toeplitz projection, $P_{\boldsymbol{\mathcal{T}}}(\mathbf{T})$, is achieved by replacing the elements along each diagonal with their mean.

\subsubsection{Projection to the Irregular Toeplitz Set}
Recall the definition of $\boldsymbol{\mathcal{T}}_{\subVecgamma}^K$ from \eqref{eq:setA_rankK}. Consider a matrix $\mathbf{T} \notin \boldsymbol{\mathcal{T}}_{\subVecgamma}^K$. The goal is to project $\mathbf{T}$ onto $\boldsymbol{\mathcal{T}}_{\subVecgamma}^K$,
\begin{equation}\label{eq:PTgamma}
P_{{\boldsymbol{\mathcal{T}}_{\subVecgamma}^K}}(\mathbf{T}) = \mathbf{W}(\boldsymbol{\gamma},\tilde{\boldsymbol{z}})\mathbf{D}\mathbf{W}(\boldsymbol{\gamma},\tilde{\boldsymbol{z}})^{\mathsf{H}},
\end{equation}
for some set of parameters $\tilde{\boldsymbol{z}}\in \mathbb{C}^{K}$. A projection onto $\boldsymbol{\mathcal{T}}_{\subVecgamma}^K$ can be achieved by constructing an irregular Toeplitz matrix using approximate parameters $\tilde{\boldsymbol{z}}$ retrieved from the local minima of the irregular null spectrum evaluated on the unit circle,
\begin{equation}\label{eq:localmin}
\tilde{\boldsymbol{z}} = \mathrm{arg} \min_{|z| = 1}^{k} \tilde{D}(z), \quad k = 1,\dots, K,
\end{equation}
where $\mathrm{arg}\min_{z}^k$ denotes the argument, $z$, which produces the $k$th smallest local minima. 

Because $\mathbf{T} \notin \boldsymbol{\mathcal{T}}_{\subVecgamma}$, the roots of its null spectrum are not on the unit circle. Instead,  approximate $\boldsymbol{z}$ parameters are estimated from the irregular null spectrum. Once $\tilde{\boldsymbol{z}}$ is known, the corresponding $\mathbf{D}$ matrix containing the signal powers is estimated \eqref{eq:lininv}. Pseudocode for projection onto $\boldsymbol{\mathcal{T}}_{\subVecgamma}^K$ is given in Algorithm \ref{alger:PT}. The algorithm involves computing the IVD for a given matrix and reconstructing the matrix from the IVD outputs. The structure of $\tilde{\mathbf{T}}$ makes clear that $\tilde{\mathbf{T}}\in\boldsymbol{\mathcal{T}}_{\subVecgamma}^K$.

\begin{algorithm}[t]
\caption{: $\tilde{\mathbf{T}} = \mathtt{P_{\boldsymbol{\mathcal{T}}_{\boldsymbol{\gamma}}}}(\mathbf{T},\boldsymbol{\gamma},K)$}
\begin{algorithmic} 
\REQUIRE $\mathbf{T} \in \mathbb{C}^{M\times M}$,\: $\boldsymbol{\gamma}\in\mathbb{R}^{M}$, $K \in\mathbb{Z}, \: M\geq K \geq 1$
\STATE $(\boldsymbol{z},\mathbf{c}) = \mathtt{IVD}(\mathbf{T},\boldsymbol{\gamma},K) $
\STATE $\tilde{\mathbf{T}} = \mathbf{W}(\boldsymbol{\gamma},\boldsymbol{z})\mathrm{diag}(\mathbf{c})\mathbf{W}(\boldsymbol{\gamma},\boldsymbol{z})^{\mathsf{H}} $
\end{algorithmic}\label{alger:PT}
\end{algorithm}

\subsubsection{Projection to the Positive Semi-Definite Cone}

The set of PSD matrices is 
\begin{equation}\label{eq:PSDset}
\mathcal{S}_{\succeq 0} = \{ \mathbf{M}\in \mathbb{C}^{N\times N} :  \lambda_i\in \mathbb{R}, \:\:  \lambda_i\geq 0, \quad \forall \: i \},
\end{equation}
where $\lambda_i$ is the $i$th eigenvalue of $\mathbf{M}$, with corresponding eigen vector $\mathbf{e}_i$. The projection onto $\mathcal{S}_{\succeq 0}$ is \cite{boyd2003alternating}
\begin{equation}\label{eq:pC}
P_{\mathcal{S}_{\succeq 0}}(\mathbf{M}) = \sum_{i=1}^{N} \max(0,\lambda_i) \mathbf{e}_i \mathbf{e}_i^\mathsf{H}.
\end{equation} 
The set $\mathcal{S}_{\succeq 0}$ forms a cone and is convex \cite{boyd2004convex}.

\subsection{Alternating Projections for NUA Gridless DOA}\label{sec:PWGD}

The AP algorithm is an optimization scheme which has found notable success when applied to the structured low rank matrix completion problem \cite{condat2015cadzow, cai2015rop, candes2015phase, cho2016fast}. The basic concept of AP is that a solution located at the intersection of two or more sets can be found by iteratively projecting an estimate between the sets. The algorithm is guaranteed to converge when all sets are convex, but convergence is not guaranteed when one or more sets is non-convex (i.e. the rank constrained set) \cite{chu2003structured}. AP can be used to solve \eqref{eq:gridlessarb}.

Define the set
\begin{equation}\label{eq:set3}
\mathcal{S}_{(\boldsymbol{\mathcal{T}}_{\subVecgamma}^K,\mathbf{Y})} = \{ \mathbf{M}: \mathbf{M} = \begin{bmatrix} 
\mathbf{T} & \mathbf{Y} \\
\mathbf{Y}^{\mathsf{H}} & \mathbf{Q} 
\end{bmatrix}, \mathbf{T} \in \boldsymbol{\mathcal{T}}_{\subVecgamma}^K \},
\end{equation}
where $\boldsymbol{\gamma}$, $\mathbf{Y}\in \mathbb{C}^{M\times L}$, and $K$ are known and $\mathbf{Q} \in \mathbb{C}^{L\times L}$ is a free matrix. Here $\boldsymbol{\gamma}$ is the array element position vector, and $\mathbf{Y}$ is the measurement matrix. The projection onto $\mathcal{S}_{(\boldsymbol{\mathcal{T}}_{\boldsymbol{\gamma}}^K,\mathbf{Y})}$ is performed as
\begin{equation}
\begin{aligned}
\mathbf{M} &= \begin{bmatrix}
\mathbf{B} &\mathbf{C} \\
\mathbf{D} &\mathbf{Q} \end{bmatrix}, \\[7pt]
P_{\mathcal{S}_{(\boldsymbol{\mathcal{T}}_{\boldsymbol{\gamma}}^K,\mathbf{Y})} }(\mathbf{M}) &= \begin{bmatrix}
 P_{\boldsymbol{\mathcal{T}}_{\boldsymbol{\gamma}}^K}(\mathbf{B}) &\mathbf{Y} \\
\mathbf{Y}^{\mathsf{H}}&\mathbf{Q} \end{bmatrix}.
\end{aligned}
\end{equation}
In words, the projection is achieved by replacing the top left submatrix of $\mathbf{M}$ with its projection to $\boldsymbol{\mathcal{T}}_{\subVecgamma}^K$ (Algorithm \ref{alger:PT}), and replacing the corner submatrices with $\mathbf{Y}$ and $\mathbf{Y}^{\mathsf{H}}$.

The AP algorithm for solving \eqref{eq:gridlessarb} is achieved by projecting an initial estimate between $\mathcal{S}_{\succeq 0}$ and $\mathcal{S}_{(\boldsymbol{\mathcal{T}}_{\subVecgamma}^K, \mathbf{Y})}$,
\begin{equation}\label{eq:AP}
\begin{split}
{\mathbf{H}}^{(i+1)} &= P_{\mathcal{S}_{\succeq 0}}(\mathbf{L}^{(i)}), \\
\mathbf{L}^{(i+1)}&= P_{\mathcal{S}_{(\boldsymbol{\mathcal{T}}_{\subVecgamma}^K, \mathbf{Y})}}(\mathbf{H}^{(i+1)}),
\end{split}
\end{equation}
for arbitrary initial estimate $\mathbf{L}^{(0)}\in \mathbb{C}^{(M+L)\times(M+L)}$. Upon convergence, the IVD of the $M\times M$ upper right submatrix of $\mathbf{S}$ retrieves the DOAs. Pseudocode for AP based gridless DOA algorithm is given in Algorithm \ref{alger:eAPG}. We deem this algorithm, ``\textit{AP gridless}".

AP gridless requires prior knowledge of $K$. There have been many works on source number estimation \cite{aouada2003comparative,han2013improved} which can be used to estimate $K$. Alternatively, AP gridless can be applied for a range of $K$, and the sparsest solution which adequately reconstructs the measurements can be taken as optimal.

\begin{figure}[t]
\begin{minipage}[t]{\linewidth}
\centering
\centerline{\includegraphics[width=\textwidth]{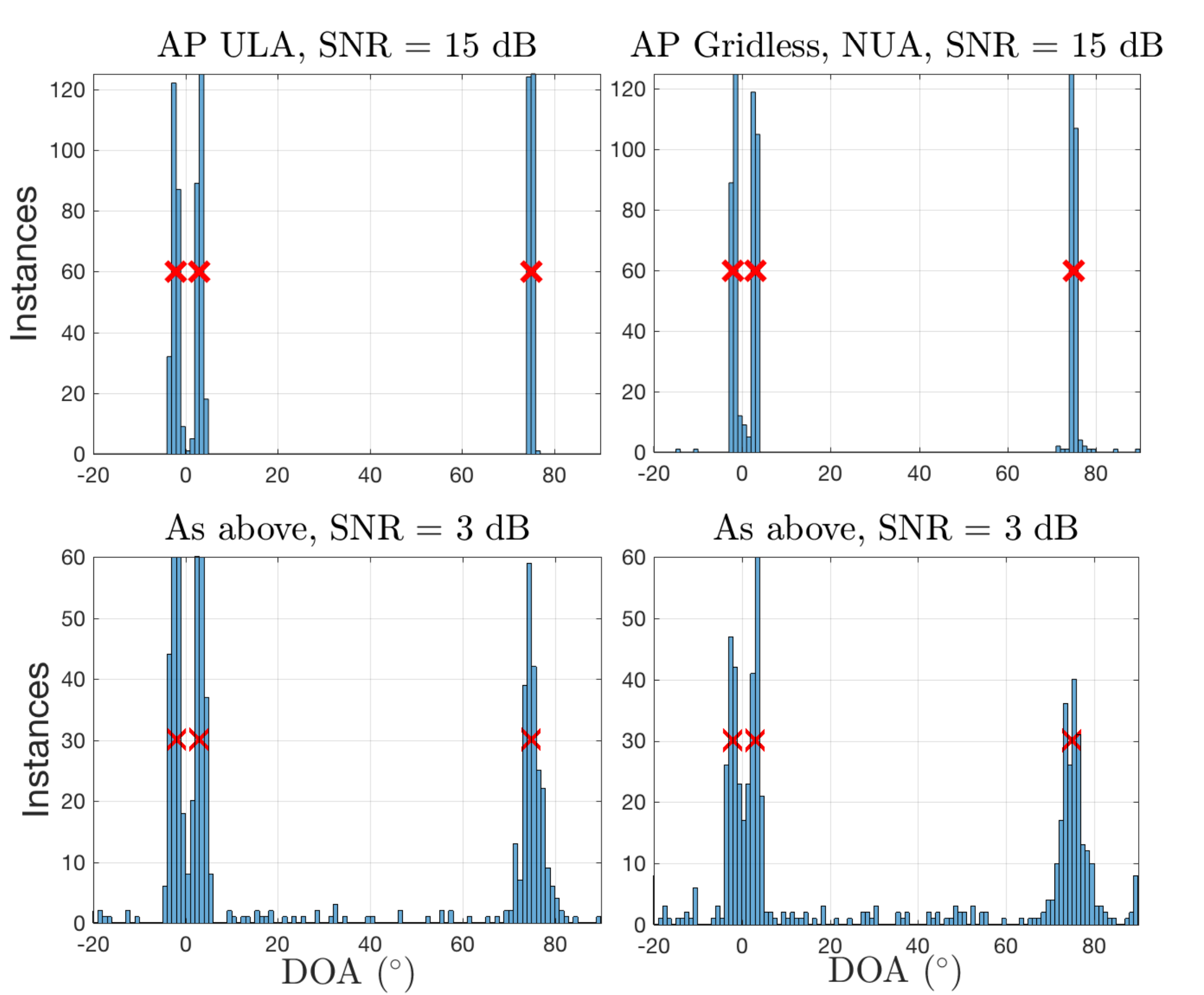}}
\caption{Histogram of recovered DOAs for AP ULA (Algorithm \ref{alger:APULA}) and AP gridless (Algorithm \ref{alger:eAPG}), $M = 20$, $L = 1$, $\sigma_s = 5$, $\boldsymbol{\theta} = [-2,3,75]^{\circ}$ (red x). 250 trials per histogram. Left- AP ULA. Right- AP Gridless (NUA).} 
\label{fig:hist}
\end{minipage}
\end{figure}

The algorithm is called ``\textit{AP ULA}" when the irregular Toeplitz projection, $P_{\boldsymbol{\mathcal{T}}_{\boldsymbol{\gamma}}^K}$ (Algorithm \ref{alger:PT}), is replaced with a Toeplitz projection, $P_{\boldsymbol{\mathcal{T}}}$ \eqref{eq:PToep}. AP ULA can only be used for measurements from a ULA, and is given in Algorithm \ref{alger:APULA}.

\begin{algorithm}[t]
\caption{: $\hat{\boldsymbol{\theta}} = \texttt{AP\_Gridless}(\mathbf{Y},\boldsymbol{\gamma},K)$}
\begin{algorithmic} 
\STATE $[M,L]$ = $\mathtt{size}(\mathbf{Y})$
\REQUIRE $\boldsymbol{\gamma} \in \mathbb{R}^{M}$, $K \in \mathbb{Z}$, $0 < K < M$
\STATE $\mathbf{L}^{(0)}$ = $[ \mathbf{0}, \mathbf{Y}; \mathbf{Y}^{\mathsf{H}}, \mathbf{I}]$
\FOR{ $i = 1:\mathtt{max\_iterations}$} 
\STATE $\mathbf{H}^{(i)}$ = $P_{\mathcal{S}_{\succeq 0}}(\mathbf{L}^{(i-1)})$
\STATE $\mathbf{T}$ = $\mathbf{H}^{(i)}(1:M, 1:M)$
\STATE $\mathbf{Q}$ = $\mathbf{H}^{(i)}(M+1:\mathtt{end}, M+1:\mathtt{end})$
\STATE $\mathbf{L}^{(i)}$ = $[\mathtt{P_{\boldsymbol{\mathcal{T}}_{\subVecgamma}}}(\mathbf{T},\boldsymbol{\gamma},K), \mathbf{Y}; \mathbf{Y}^{\mathsf{H}}, \mathbf{Q}]$
\IF{$\|\mathbf{L}^{(i)}-\mathbf{L}^{(i-1)}\|_F \leq \mathtt{1e-7}$ }
\STATE \texttt{break}
\ENDIF
\ENDFOR
\STATE $\mathbf{T}$ = ${\mathbf{L}}(1:M, 1:M)$
\STATE $[\boldsymbol{z},\sim]$ = $\mathtt{IVD}(\mathbf{T},\boldsymbol{\gamma},K)$
\STATE $\hat{\boldsymbol{\theta}}  = -\mathtt{asin}(\mathtt{angle}({\boldsymbol{z}})/\pi)$
\end{algorithmic}\label{alger:eAPG}
\end{algorithm}

 \subsection{Initialization and Convergence}

The AP algorithm is guaranteed to converge at a linear rate when applied between two closed convex sets \cite{bauschke1993convergence}, however the set of matrices with rank $\leq K$ is non-convex, (rather, it is quasi-convex \cite{dattorro2010convex}). It is not known if AP will converge when applied between a convex set and the rank constrained set. By construction, $\boldsymbol{\mathcal{T}}_{\subVecgamma}^K$ is the intersection between $\boldsymbol{\mathcal{T}}_{\subVecgamma}$ and the rank constrained set, thus it is also non-convex and convergence remains an open question.

Progress towards a convergence proof was achieved in \cite[Theorem 3.2]{attouch2010proximal} and can be summarized as it applies to \eqref{eq:AP} as follows:
\vskip .1in
\textbf{Theorem:} \textit{Let $\mathbf{L}^{(i)}, \mathbf{H}^{(i)}$ be generated according to \eqref{eq:AP}, and $\mathbf{L}^{(0)}\in \mathcal{S}_{(\boldsymbol{\mathcal{T}}_{\subVecgamma}^K,\mathbf{Y})}$ then}
\vskip .05in
\begin{enumerate}
 \item{Either $\| \mathbf{L}^{(i)} - \mathbf{H}^{(i)} \|_F^2 \rightarrow \infty$ as $i \rightarrow \infty$ or $\|\mathbf{L}^{(i)} - \mathbf{H}^{(i)}\|_{F}^2$ converges to 0.}
\vskip .05in 
\item{If $\mathbf{L}^{(0)}$ is sufficiently close to the global minimizer of \eqref{eq:gridlessarb}, $\mathbf{L}^{(0)}$ converges to the global minimum.}
\vskip .1in
\end{enumerate}

Note that Theorem 1 does not guarantee the point of convergence will be the global minimizer unless the initialization point is sufficiently nearby the optimal solution. In general, it can only be assumed the convergence point will be a local minimum of the optimization function. 

\begin{figure}[t]
\begin{minipage}[t]{\linewidth}
\centering
\centerline{\includegraphics[width=3.5in]{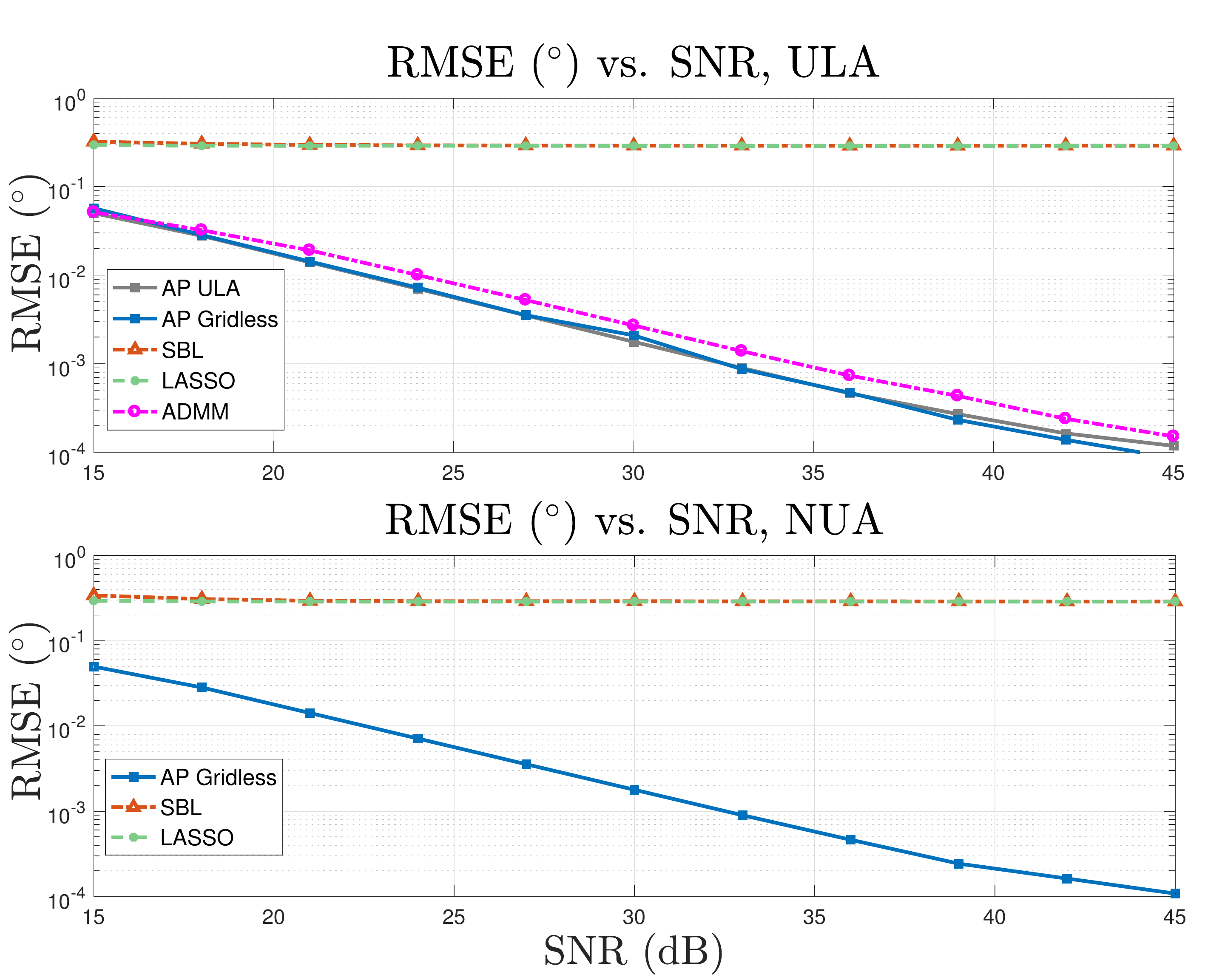}}
\caption{$\mathrm{RMSE}$ vs. SNR on high SNR single snapshot measurements using AP Gridless (Algorithm \ref{alger:eAPG}), AP ULA (Algorithm \ref{alger:APULA}), ADMM (App. \ref{sec:Appendix}, $\tau = 10^{-5}$), SBL, and LASSO. $M = 20, K = 3, L = 1, \sigma_s = 5$. Each point represents average over 250 trials. All algorithms run to convergence.}
\label{fig:highSNR}
\end{minipage}
\end{figure}

\begin{algorithm}[t]
\caption{: $\hat{\boldsymbol{\theta}}$ = \texttt{AP\_ULA}($\mathbf{Y},\boldsymbol{\gamma},K)$}
\begin{algorithmic} 
\STATE $[M,L]$ = $\mathtt{size}(\mathbf{Y})$
\REQUIRE $\boldsymbol{\gamma} \in \mathbb{R}^{M}$, $K \in \mathbb{Z}$, $0 < K < M$
\STATE $\mathbf{L}^{(0)}$ = $[ \mathbf{0}, \mathbf{Y}; \mathbf{Y}^{\mathsf{H}}, \mathbf{I}]$
\FOR{ $i = 1:\mathtt{max\_iterations}$} 
\STATE $\mathbf{H}^{(i)}$ = $P_{\mathcal{S}_{\succeq 0}}(\mathbf{L}^{(i-1)})$
\STATE $\mathbf{T}$ = $\mathbf{H}^{(i)}(1:M, 1:M)$
\STATE $\mathbf{Q}$ = $\mathbf{H}^{(i)}(M+1:\mathtt{end}, M+1:\mathtt{end})$
\STATE $\mathbf{L}^{(i)}$ = $[P_{\boldsymbol{\mathcal{T}}}(\mathbf{T}), \mathbf{Y}; \mathbf{Y}^{\mathsf{H}}, \mathbf{Q}]$
\IF{$\|\mathbf{L}^{(i)}-\mathbf{L}^{(i-1)}\|_F \leq \mathtt{1e-7}$ }
\STATE \texttt{break}
\ENDIF
\ENDFOR
\STATE $\mathbf{T}$ = ${\mathbf{L}}(1:M, 1:M)$
\STATE $[\boldsymbol{z},\sim]$ = $\mathtt{IVD}(\mathbf{T},\boldsymbol{\gamma},K)$
\STATE $\hat{\boldsymbol{\theta}}  = -\mathtt{asin}(\mathtt{angle}({\boldsymbol{z}})/\pi)$
\end{algorithmic}\label{alger:APULA}
\end{algorithm}

All simulations in Sec. \ref{sec:Sim} were performed using the initialization,
\begin{equation}
\mathbf{L}^{(0)} = \begin{bmatrix}
\mathbf{0}\: &\mathbf{Y} \\
\mathbf{Y}^{\mathsf{H}} & \mathbf{I} \end{bmatrix},
\end{equation}
which was always observed to converge to a critical point rather than diverging. Because it remains unknown if the initialization point will bring the algorithm to a local or global minimum, all further analysis on the accuracy of AP for gridless DOA is left to numerical simulation performed in Sec. \ref{sec:Sim}.

\section{Simulation}\label{sec:Sim}

AP gridless  \eqref{eq:AP} (Algorithm \ref{alger:eAPG}), and AP ULA \eqref{eq:PToep} (Algorithm \ref{alger:APULA}) were applied to solve the gridless DOA problem \eqref{eq:gridlessarb} for simulated measurements modeled by \eqref{eq:model}. For each simulation, $K$ randomly generated DOAs were chosen such that the minimum angular separation was  $\frac{1}{M}$ for DOAs randomly drawn between $[0,1)$ and scaled to $[-90^{\circ},90^{\circ})$. Signals contained in the $K$ rows of $\mathbf{X}$ were generated as complex values with uniformly distributed phase. The signal from each DOA was given random amplitude $\sigma_s^x, \: x\in\mathcal{U}(0,1)$ to model sources with different powers. Noise, $\mathbf{N}$, was drawn from a complex Gaussian distribution with mean zero and variance $\mathbf{I}$, then scaled to fit the desired signal to noise ratio (SNR) determined by
\begin{equation}\label{eq:SNR}
\mathrm{SNR} = 10\log_{10}\Big(\frac{\| \mathbf{Z} \|_F^2}{\mathbf{\| N} \|_F^2}\Big).
\end{equation}
Array measurements were simulated using an array of $M$ sensors distributed as both a ULA and NUA. The NUA geometry was generated by adding random offsets drawn from a uniform distribution between $[-.5, .5)$ (in units of half-wavelengths) to each sensor position. No restriction was placed on how nearby two NUA elements could be.  

\begin{figure}[t]
\begin{minipage}[t]{\linewidth}
\centering
\centerline{\includegraphics[width=\textwidth]{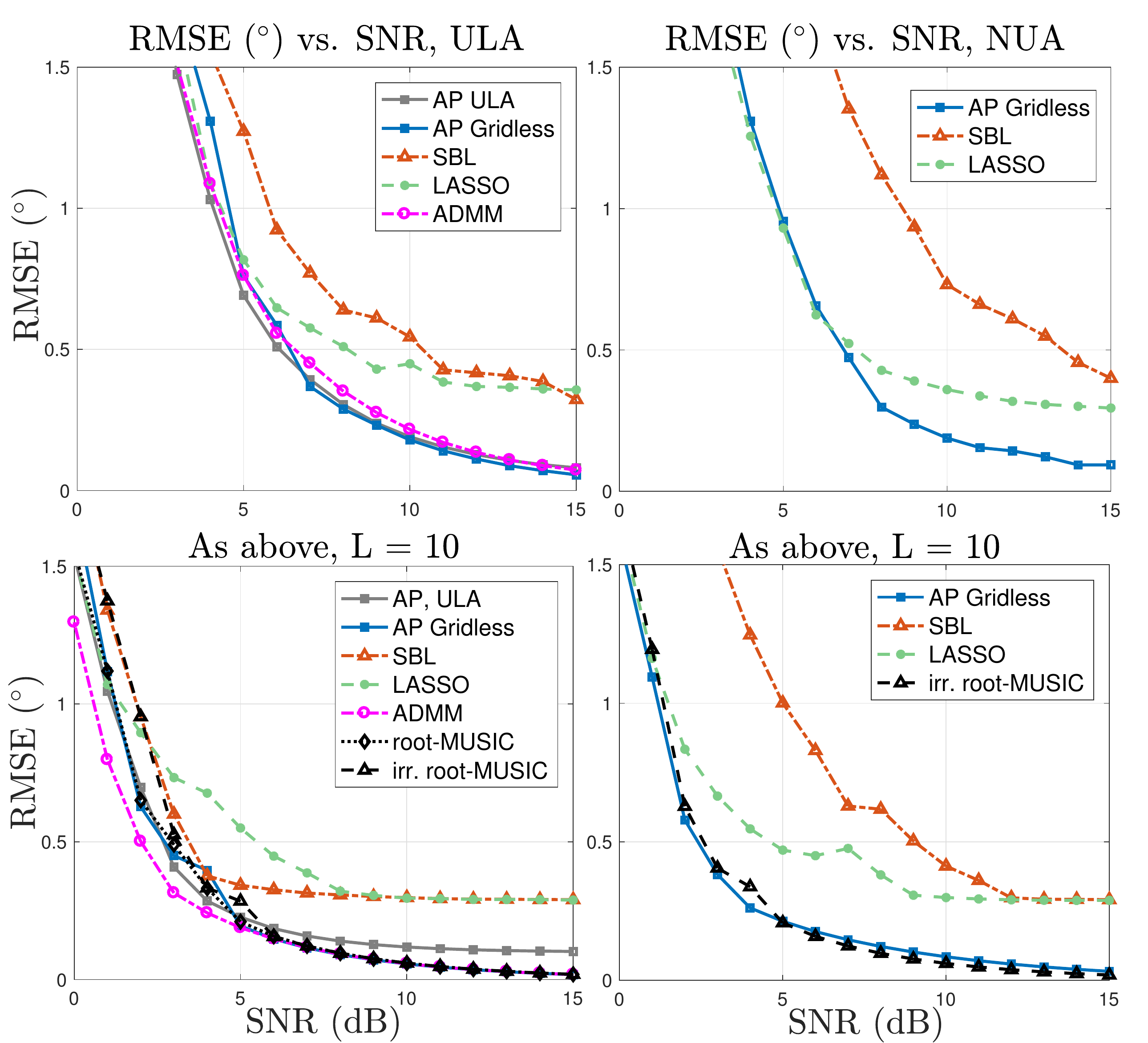}}
\caption{ RMSE vs. SNR for AP gridless (Algorithm \ref{alger:eAPG}), AP ULA (Algorithm \ref{alger:APULA}), ADMM (App. \ref{sec:Appendix}), SBL, LASSO, root-MUSIC, and irregular root-MUSIC (Algorithm \ref{alger:eRM}), $M = 20$, $K = 3$, $\sigma_s = 5$. Each point represents 250 trials. Top left- ULA measurements, $L = 1$. Top right- NUA measurements, $L = 1$. Bottom left- ULA measurements, $L = 10$. Bottom right- NUA measurements, $L = 10$.}
\label{fig:SNR}
\end{minipage}
\end{figure}

All simulations were performed for ULA and NUA. In the ULA case, AP gridless and AP ULA were both applied. It was assumed $K$ was known for each simulation. AP gridless and AP ULA were considered converged when $\|\mathbf{L}^{(i)}-\mathbf{L}^{(i-1)}\|_F \leq 10^{-7}$, or terminated after $10 K$ iterations. 

The accuracy of the solution was gauged by the root mean square error (RMSE) between true, $\theta_k$, and recovered, $\hat{\theta}_k$, DOAs defined as
\begin{equation}
\mathrm{RMSE} = \sqrt{ \mathbb{E} \Bigg[\frac{1}{K}\sum_{k=1}^{K} (\theta_k - \hat{\theta}_k)^2} \Bigg].
\end{equation}
%
A maximum MSE threshold of $10^{\circ}$ was used to provide an even error penalty for simulations resulting in incorrect DOA estimates.

Sparse Bayesian learning (SBL) \cite{wipf2004sparse,yang2012off, gerstoft2016multisnapshot, gemba2017multi,gemba2019robust}, and least absolute shrinkage and selector operator (LASSO) \cite{tibshirani1996regression,edelmann2011beamforming,xenaki2014compressive,gerstoft2015multiple} methods were given a dictionary of the array manifold with $1^{\circ}$ separation between entries (180 total). The LASSO tuning parameter was selected such that the solution was $K$ sparse (which requires knowledge of $\| \mathbf{N} \|_F$ \cite{malioutov2005sparse}). The ADMM algorithm was applied to ULA measurements using parameters $\tau = .01$, and $\rho = 1$ (see App. \ref{sec:Appendix}). 



\begin{figure}[t]
\begin{minipage}[t]{\linewidth}
\centering
\centerline{\includegraphics[width=\textwidth]{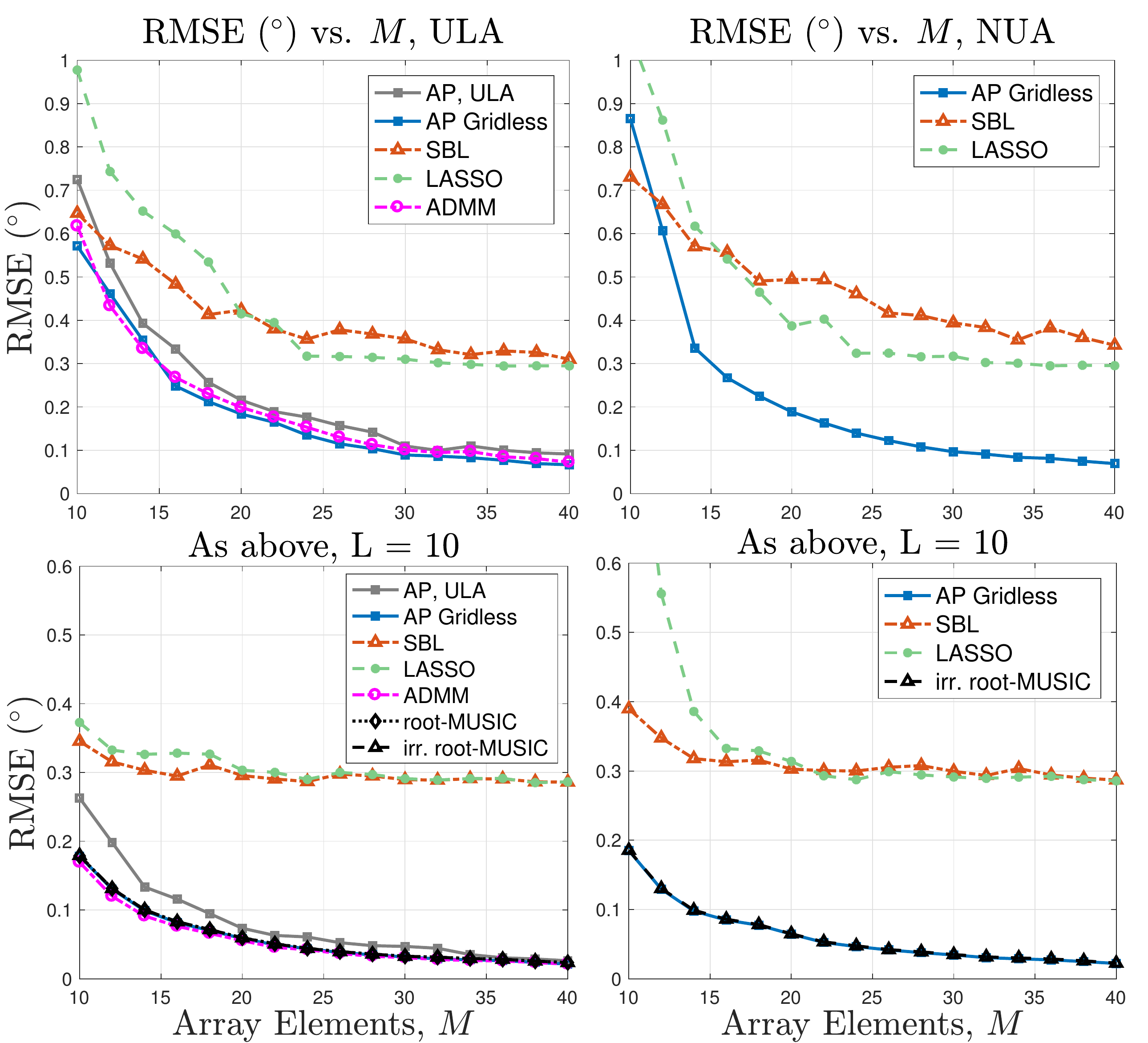}}
\caption{ RMSE vs. $M$ for AP gridless (Algorithm \ref{alger:eAPG}), AP ULA (Algorithm \ref{alger:APULA}), ADMM \ref{sec:Appendix}, SBL, LASSO, root-MUSIC, and irregular root-MUSIC (Algorithm \ref{alger:eRM}), $\mathrm{SNR} = 10$ dB, $K = 3$, $\sigma_s = 5$. Each point represents 250 trials. Top left- ULA measurements, $L = 1$. Top right- NUA measurements, $L = 1$. Bottom left- ULA measurements, $L = 10$. Bottom right- NUA measurements, $L = 10$.}
\label{fig:M}
\end{minipage}
\end{figure}

Example results from AP ULA and AP gridless are detailed in Fig. \ref{fig:hist}. The AP ULA algorithm never misclassifies a DOA when the measurement SNR is high (Fig. \ref{fig:hist}, top left), and lowering the measurement SNR results in few misclassifications (Fig. \ref{fig:hist}, bottom left). In contrast, there are some NUA geometries which cause AP gridless to misclassify a DOA, even for high SNR (Fig. \ref{fig:hist}, top right). The rate of misclassification for the low SNR NUA case is slightly higher than that of the ULA case (Fig. \ref{fig:hist}, bottom right). 

Gridless DOA excels over gridded methods in the high SNR, low snapshot case. This is due to quantization error from grid mismatch limiting the maximum accuracy of gridded techniques. Figure \ref{fig:highSNR} compares gridless and gridded algorithms for high SNR, single snapshot measurements when all algorithms are run to convergence. The best achievable accuracy for a gridded technique using $1^{\circ}$ DOA separation is RMSE = $.25^{\circ}$. In the ULA case (Fig. \ref{fig:highSNR}, top), all gridless techniques achieve accuracy proportional to SNR. ADMM is not as precise due to limitations in parameter tuning. In the NUA case (Fig. \ref{fig:highSNR}, bottom) AP gridless maintains excellent performance. 

\begin{figure}[t]
\begin{minipage}[t]{\linewidth}
\centering
\centerline{\includegraphics[width=\textwidth]{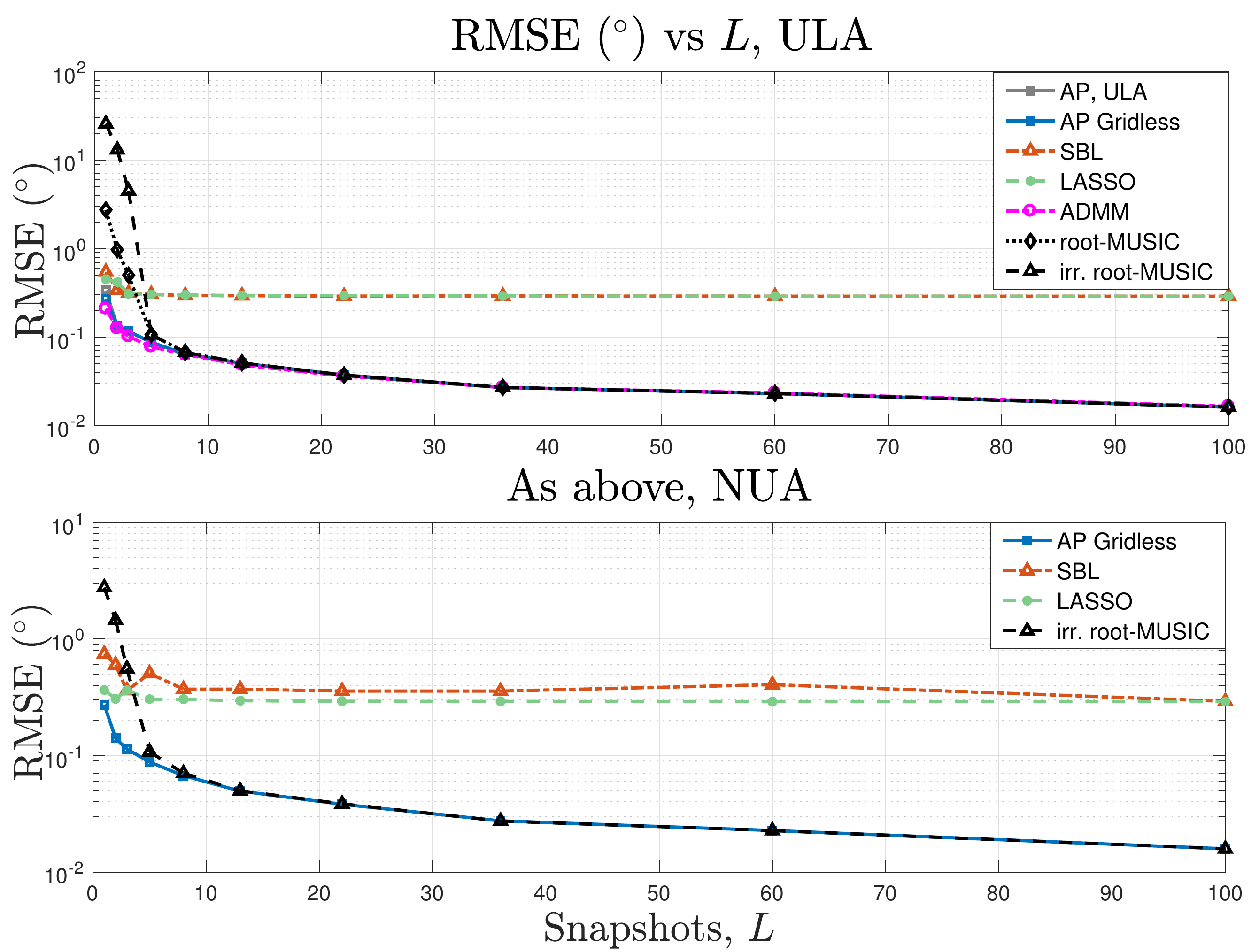}}
\caption{ RMSE vs. $L$ for AP gridless (Algorithm \ref{alger:eAPG}), AP ULA (Algorithm \ref{alger:APULA}), ADMM (App. \ref{sec:Appendix}), SBL, LASSO, root-MUSIC, and irregular root-MUSIC (Algorithm \ref{alger:eRM}), $M = 20$, $K = 3$, $\sigma_s = 5$, $\mathrm{SNR} = 10$ dB. Each point represents 250 trials. Top- ULA measurements. Bottom- NUA measurements.}
\label{fig:L}
\end{minipage}
\end{figure}

The performance of each algorithm in lower SNR scenarios is detailed in Fig. \ref{fig:SNR}. In the ULA case, AP gridless and AP ULA achieve similar performance to ADMM, particularly in the single snapshot case (Fig. \ref{fig:SNR}, top left), indicating that convex relaxation is unnecessary to solve rank minimization problems. AP gridless and AP ULA perform slightly worse than ADMM for low SNR multi-snapshot measurements (Fig. \ref{fig:SNR}, bottom left), suggesting that ADMM is more robust to noise. AP gridless is superior to gridded techniques for NUA measurements (Fig. \ref{fig:SNR}, top and bottom right), except in low SNR scenarios where LASSO can attain similar performance. Irregular root MUSIC (Algorithm \ref{alger:eRM}) attains near identical performance to AP gridless in the multiple snapshot NUA case (Fig. \ref{fig:SNR}, bottom right), for less computational cost.


Performance of the algorithms versus array elements, $M$, is given in Fig. \ref{fig:M}. AP Gridless and ADMM are superior in the ULA case (Fig. \ref{fig:M}, top left), as gridded methods are limited in accuracy by grid resolution. For NUAs, AP gridless and irregular root-MUSIC are superior for the same reason. Root-MUSIC and irregular root-MUSIC attain equal performance to AP gridless and ADMM in the multi-snapshot case (Fig. \ref{fig:M}, bottom left and right) for lower computational complexity. 
This is also shown by Fig. \ref{fig:L}, which compares each method vs. number of snapshots, $L$. Once the number of snapshots is greater than the number of sources ($L\geq K$), there is no benefit to choosing a gridless technique because root-MUSIC and irregular root-MUSIC attain the same performance as gridless methods for reduced computational cost.

\begin{figure}[t]
\begin{minipage}[t]{\linewidth}
\centering
\centerline{\includegraphics[width=\textwidth]{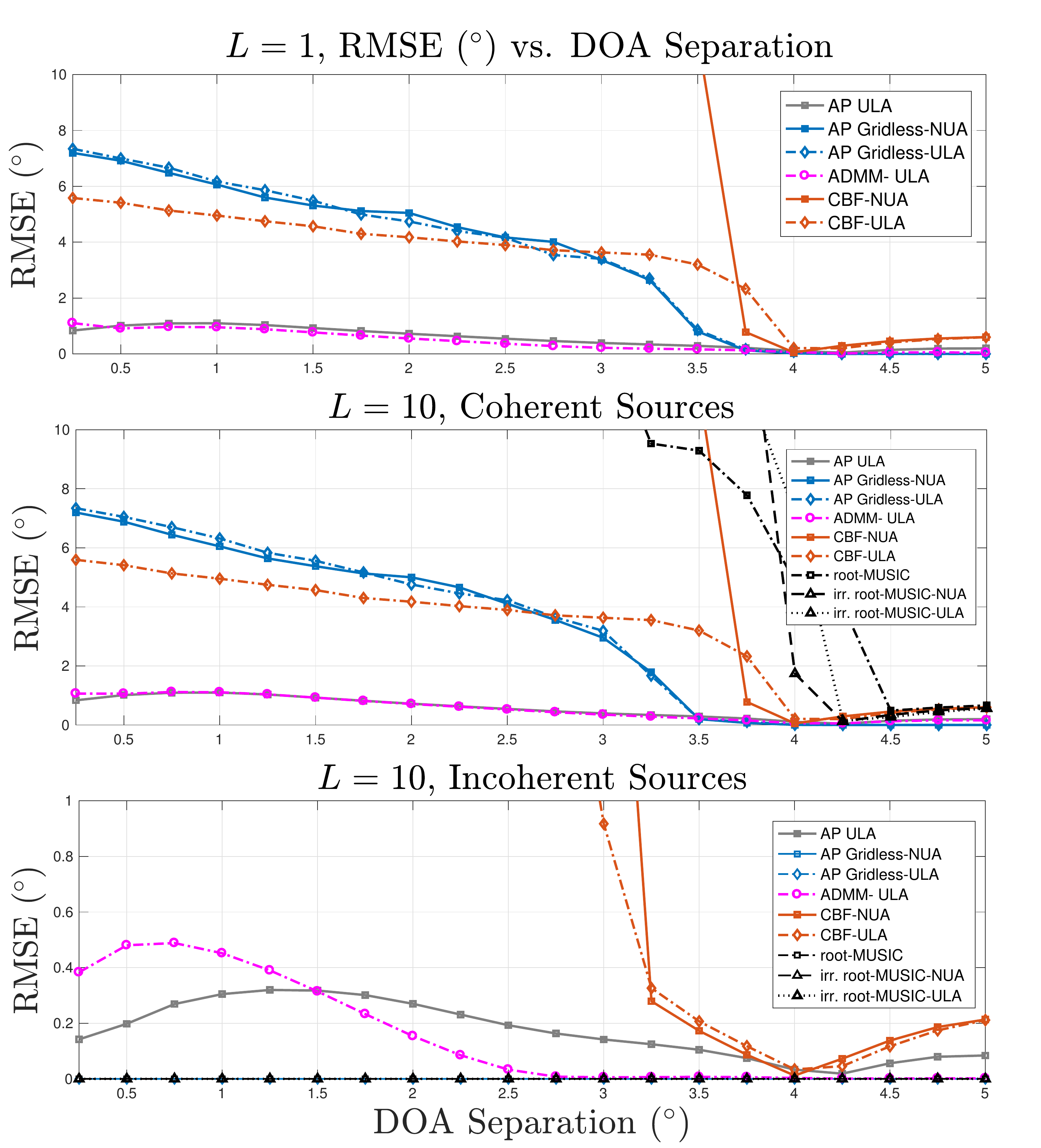}}
\caption{$\mathrm{RMSE}$ vs. DOA separation of conventional (Bartlett) beamformer (CBF), AP gridless (Algorithm \ref{alger:eAPG}), AP ULA (Algorithm \ref{alger:APULA}), ADMM (App. \ref{sec:Appendix}), root-MUSIC, and irregular root-MUSIC (Algorithm \ref{alger:eRM}) on noiseless simulated ULA and NUA measurements using $K = 2$ equal amplitude sources at $\pm\theta^{\circ}$, $M=20$. No maximum MSE penalty, each point represents 250 trials. Top- $L = 1$. Middle- $L = 10$, coherent sources. Bottom- $L = 10$, incoherent sources.}
\label{fig:sep}
\end{minipage}
\end{figure}

To get an indication of the high resolution capability of each algorithm, Fig. \ref{fig:sep} compares AP gridless and AP ULA to conventional beamforming (CBF) and root-MUSIC for noiseless measurements with $2$ DOAs at $\pm\theta^{\circ}$. For the single snapshot case (Fig. \ref{fig:sep}, top), AP gridless has difficulty resolving nearby DOAs while ADMM and AP ULA have outstanding performance. The difference is that ADMM and AP ULA use projection to $\boldsymbol{\mathcal{T}}$ in place of projection to $\boldsymbol{\mathcal{T}}_{\subVecgamma}^K$. 

$P_{\boldsymbol{\mathcal{T}}_{\subVecgamma}^K}$ \eqref{eq:PTgamma}, (Algorithm \ref{alger:PT}) is fundamentally different than $P_{\boldsymbol{\mathcal{T}}}$ \eqref{eq:PToep}. ${P}_{\boldsymbol{\mathcal{T}}}$ outputs the nearest Toeplitz matrix in Frobenius norm to its input (orthogonal projection), which is generally a full rank Toeplitz matrix. In contrast, $P_{\boldsymbol{\mathcal{T}}_{\subVecgamma}^K}$ outputs a rank $K$ matrix which is not necessarily the orthogonal projection to ${\boldsymbol{\mathcal{T}}_{\subVecgamma}^K}$. This is because $P_{\boldsymbol{\mathcal{T}}_{\subVecgamma}^K}$ is estimated by reconstructing an irregular Toeplitz matrix using $\boldsymbol{z}$ parameters given by the local minima of $\tilde{D}(z)\rvert_{|z| = 1}$, rather than the phase angle of the roots of $\tilde{D}(z)$. Roots of $\tilde{D}(z)$ which are nearby in phase angle sometimes do not produce unique local minima of $\tilde{D}(z)\rvert_{|z| = 1}$, resulting in suboptimal $\boldsymbol{z}$ parameter estimates. This is only an issue when the DOAs are poorly separated and the measurements do not give a precise estimate of the noise subspace ($L < K$).


The same problem exists in the multiple snapshot ($L\geq K$) coherent sources case (Fig. \ref{fig:sep}, middle). Because the sources are coherent, the noise subspace still cannot be well estimated and AP gridless fails to resolve nearby DOAs until they are sufficiently separated. 

When the sources are made incoherent (Fig. \ref{fig:sep}, bottom), the noise subspace can be estimated precisely, and AP gridless, root-MUSIC, and irregular root-MUSIC are exact. In contrast, ADMM and AP ULA are not as precise because $P_{\boldsymbol{\mathcal{T}}}$ does not project to a specifically rank $K$ solution, resulting in `noisy' DOA estimates.

The computational bottleneck of AP gridless is the projection to the irregular Toeplitz set, $P_{\boldsymbol{\mathcal{T}}_{\subVecgamma}^K}$, where a spectral search over $\tilde{D}(z)\rvert_{|z|=1}$ is executed. In contrast, the AP ULA and ADMM algorithms use the relatively efficient projection to the Toeplitz set, $P_{\boldsymbol{\mathcal{T}}}$ \eqref{eq:PToep}, and are limited by the eigen-decomposition of the $\mathbf{S}$ matrix in its projection to the positive semi-definite cone, $P_{\mathcal{S}\succeq0}$. Both algorithms are relatively fast, having runtime under one second on a regular CPU for problems with $(M+L) < 100$.

\section{Conclusion}

The problem of gridless direction of arrival (DOA) estimation for non-uniform array (NUA) geometries was considered. Towards this goal, the irregular Vandermonde decomposition (IVD) was introduced, which is a generalization of the Vandermonde decomposition for irregularly sampled signals, such as those sampled from a NUA. From the perspective of the IVD, any covariance matrix can be seen as an irregular Toeplitz matrix, and can be decomposed back into its irregular Vandermonde components. The decomposition can be used to extend gridless DOA to NUAs, as well as extending the larger continuous compressed sensing (CCS) problem to irregularly sampled signals. The decomposition also extends root-MUSIC to NUAs.

An intuitive non-convex method of solving gridless DOA for NUAs based on the alternative projections (AP) algorithm was proposed. The proposed algorithm was named AP gridless. Simulation on uniform linear array (ULA) measurements found AP gridless attains similar performance to its convex counterpart, ADMM. For NUAs, ADMM cannot be generalized, and only AP gridless applies. AP gridless was found to be robust to noise, high resolution, and has superior performance compared to grid based techniques for high SNR.


 \renewcommand{\theequation}{A-\arabic{equation}}
  \setcounter{equation}{0}  
\begin{appendices} \label{sec:ADMM}
\section{ADMM for gridless DOA}\label{sec:Appendix}
Here the alternating directions method of multipliers (ADMM) formulation of gridless DOA is detailed as it applies to multiple snapshot ULA measurements. A review of ADMM is given in \cite{boyd2011distributed}. Other sources which provide details on ADMM specifically for gridless DOA are \cite{YANG2018509, bhaskar2012atomic,  li2014off, yang2019two,semper2019admm}.

Start with the rank minimization problem of \eqref{eq:gridless2}. Because the the rank constraint is non-convex the problem must be cast to its convex relaxation before ADMM is applicable,
\begin{equation}\label{eq:convex_relaxation}
\begin{aligned}
&\min_{\mathbf{Q},\mathbf{S}, \hat{\mathbf{Y}},\mathbf{u}} \quad && \frac{1}{2}\| \hat{\mathbf{Y}}-\mathbf{Y} \|_2^2 + \frac{\tau}{2}\Big(\mathrm{Tr}(\mathbf{Q}) + \mathrm{Tr}(\boldsymbol{\mathcal{T}}(\mathbf{u}))\Big) \\
& \mathrm{subject\: to}\quad&& \mathbf{S} = \begin{bmatrix} \boldsymbol{\mathcal{T}}(\mathbf{u})  &\hat{\mathbf{Y}} \\
 \hat{\mathbf{Y}}^{\mathsf{H}} &\mathbf{Q}\end{bmatrix}, \quad \mathbf{S} \succeq 0,
\end{aligned}
\end{equation}
where $\mathbf{Y}\in\mathbb{C}^{M\times L}$ is the measurement matrix, $\boldsymbol{\mathcal{T}}(\mathbf{u})\in \mathbb{C}^{M\times M}$ is the Toeplitz matrix whose first column is $\mathbf{u}$, $\mathbf{Q}\in\mathbb{C}^{L\times L}$, and $\tau$ is a user defined parameter.

Next, the augmented Lagrangian is formed by collecting the constraints into the optimization function \cite{bhaskar2012atomic,li2014off},
\begin{multline} \label{eq:Lagrange}
\mathcal{L}_{\rho}(\mathbf{Q},\mathbf{S},\hat{\mathbf{Y}},\mathbf{\Lambda},\mathbf{u}) = \frac{1}{2} \| \hat{\mathbf{Y}}-\mathbf{Y} \|_2^2 + \frac{\tau}{2}\Big(\mathrm{Tr}(\mathbf{Q})+\mathrm{Tr}(\boldsymbol{\mathcal{T}}(\mathbf{u}))\Big) + \\
\Big\langle \mathbf{\Lambda}, \mathbf{S} - \begin{bmatrix} \boldsymbol{\mathcal{T}}(\mathbf{u}) &\hat{\mathbf{Y}} \\ 
\hat{\mathbf{Y}}^{\mathsf{H}} &\mathbf{Q} \end{bmatrix} \Big\rangle + \frac{\rho}{2}\Big\| \mathbf{S} - \begin{bmatrix} \boldsymbol{\mathcal{T}}(\mathbf{u}) &\hat{\mathbf{Y}} \\ 
\hat{\mathbf{Y}}^{\mathsf{H}} &\mathbf{Q} \end{bmatrix} \Big \|_F^2,
\end{multline}
where $\rho$ is a user defined parameter and $\langle . \,,\, .\rangle$ is the matrix inner product, and $\mathbf{\Lambda}$ is the Lagrange multiplier variable. ADMM is performed by iteratively optimizing \eqref{eq:Lagrange} over each variable independent of the the other variables. The update steps at iteration $i$ are \cite{bhaskar2012atomic,li2014off},
\begin{align}
(\mathbf{Q}^{(i+1)},&\mathbf{u}^{(i+1)},\hat{\mathbf{Y}}^{(i+1)}) \leftarrow \mathrm{arg} \: \min_{\mathbf{Q},\hat{\mathbf{Y}},\mathbf{u}} \mathcal{L}_{\rho}(\mathbf{Q},\mathbf{S}^{(i)},\hat{\mathbf{Y}}, \mathbf{\Lambda}^{(i)}, \mathbf{u}), \\
\mathbf{S}^{(i+1)} &\leftarrow \mathrm{arg} \min_{\mathbf{S}\succeq 0} \mathcal{L}_{\rho}(\mathbf{Q}^{(i+1)}, \mathbf{S}, \hat{\mathbf{Y}}^{(i+1)},\mathbf{\Lambda}^{(i)},\mathbf{u}^{(i+1)}), \\
\mathbf{\Lambda}^{(i+1)} &\leftarrow \mathbf{\Lambda}^{(i)} + \rho \Big( \mathbf{S}^{(i+1)}- \begin{bmatrix} \boldsymbol{\mathcal{T}}(\mathbf{u}^{(i+1)}) &\hat{\mathbf{Y}}^{(i+1)} \\ \hat{\mathbf{Y}}^{(i+1)\mathsf{H}} &\mathbf{Q}^{(i+1)} \end{bmatrix}
\Big).
\end{align}
The update steps for $\mathbf{Q}$, $\mathbf{u}$, and $\hat{\mathbf{Y}}$ are computable in closed form. First define the partitions
\begin{align}
\mathbf{S}^{(i)} = \begin{bmatrix} \mathbf{S}_{\boldsymbol{\mathcal{T}}} &\mathbf{S}_{\mathbf{Y}} \\ \mathbf{S}_{\mathbf{Y}^{\mathsf{H}}} &\mathbf{S}_{\mathbf{Q}}\end{bmatrix} \quad \mathrm{and} \quad \mathbf{\Lambda}^{(i)} = \begin{bmatrix} \mathbf{\Lambda}_{\boldsymbol{\mathcal{T}}} &\mathbf{\Lambda}_{\mathbf{Y}} \\ \mathbf{\Lambda}_{\mathbf{Y}^{\mathsf{H}}} &\mathbf{\Lambda}_{\mathbf{Q}}\end{bmatrix},
\end{align}
where the dimensions of each partition are the dimensions of $\boldsymbol{\mathcal{T}}, \mathbf{Y},$ and $\mathbf{Q}$. The update steps are \cite{bhaskar2012atomic,li2014off},
\begin{align}
\mathbf{Q}^{(i+1)} &= \frac{1}{2}\mathbf{S}_{\mathbf{Q}}^{(i)} + \frac{1}{2}(\mathbf{S}_{\mathbf{Q}}^{(i)})^{\mathsf{H}} + \frac{1}{\rho}\big( \mathbf{\Lambda}_{\mathbf{Q}}^{(i)} - \frac{\tau}{2}\mathbf{I}\big), \\
\mathbf{u}^{(i+1)} &= \mathcal{T}^{-1}\Big(\mathbf{S}_{\boldsymbol{\mathcal{T}}}^{(i)} + \frac{1}{\rho}\mathbf{\Lambda}_{\boldsymbol{\mathcal{T}}}^{(i)}\Big) - \frac{\tau}{2\rho}\mathbf{e}_1, \\
\hat{\mathbf{Y}}^{(i+1)} &= \frac{1}{2\rho + 1}\big( {\mathbf{Y}} + \rho\mathbf{S}_{\mathbf{Y}}^{(i)} + \rho\mathbf{S}_{\mathbf{Y}^{\mathsf{H}}}^{(i)} + 2\mathbf{\Lambda}_{\mathbf{Y}}^{(i)}\big), 
\end{align}
%
%
%
where $\mathcal{T}^{-1}$ is a function giving the first column of the nearest Toeplitz matrix to its input \eqref{eq:PToep}, and $\mathbf{e}_1$ is the canonical basis vector in $M$ dimensional space, $[1\: 0\:\dots\: 0]^{\mathsf{T}}$. 

Finally, the $\mathbf{S}$ update is a projection onto the positive semidefinite cone \eqref{eq:pC},
\begin{equation}
\mathbf{S}^{(i+1)} = P_{\mathcal{S}\succeq 0}\Big( \mathbf{S}^{(i)} - \begin{bmatrix} \boldsymbol{\mathcal{T}}(\mathbf{u}^{(i+1)}) &\hat{\mathbf{Y}}^{(i+1)} \\ \hat{\mathbf{Y}}^{(i+1)\mathsf{H}} &\mathbf{Q}^{(i+1)}\end{bmatrix} + \frac{1}{\rho}\mathbf{\Lambda}^{(i)}\Big).
\end{equation}
ADMM is known to converge at a linear rate for convex problems \cite{boyd2011distributed}.
\end{appendices}

\bibliography{gbap2}
\bibliographystyle{IEEEbib}
\end{document}